\documentclass[11pt]{article}

\usepackage{cite}
\usepackage{amsmath,amssymb,amsfonts}
\usepackage{graphicx}
\usepackage{amsthm}
\usepackage{bbm}
\usepackage{textcomp}
\usepackage{tikz}
\usepackage{pgfplots}
\usepackage{caption}
\usepackage{subcaption}
\usepackage{xcolor}
\usepackage{algorithmicx}
\usepackage[noend]{algpseudocode}
\usepackage{algorithm}
\usepackage{mathtools}
\usepackage{hyperref}
\allowdisplaybreaks 
\usepackage{algorithm,algcompatible,mathtools}
\usepackage[multidot]{grffile}
\usepackage[normalem]{ulem} 

\newcommand{\vertiii}[1]{{\vert\kern-0.25ex\vert\kern-0.25ex\vert #1\vert\kern-0.25ex\vert\kern-0.25ex\vert}}

\usepackage{array}
\usepackage{authblk}
\usepackage{setspace}
\usepackage{fullpage,etoolbox}

\newtheorem{theorem}{Theorem}
\newtheorem{lemma}{Lemma}
\newtheorem{assumption}{Assumption}
\newtheorem{proposition}{Proposition}

\newtheorem{definition}{Definition} 
\newtheorem{problem}{Problem} 

\newtoggle{draft}
\togglefalse{draft}

\DeclareMathOperator*{\argmin}{arg\!\min}

\newcommand{\RR}{\ensuremath{\mathbb{R}}}
\newcommand{\cbf}{h}

\newcommand{\statedim}{n}
\newcommand{\inputdim}{m}
\newcommand{\xx}{x}

\newcommand{\uu}{u}
\newcommand{\g}{g}

\newcommand{\f}{f}

\newcommand{\Ss}{\ensuremath{\mathcal{S}}}

\newcommand{\D}{\ensuremath{\mathcal{D}}}

\newcommand{\C}{\ensuremath{\mathcal{C}}}

\newcommand{\I}{\ensuremath{\mathcal{I}}}
\newcommand{\U}{\ensuremath{\mathcal{U}}}
\newcommand{\B}{\ensuremath{\mathcal{B}}}
\newcommand{\Lc}{L_h}

\newcommand{\norm}[1]{\| #1 \|}

\newcommand{\Zsafe}{Z_{\mathrm{safe}}}

\newcommand{\gammasafe}{\gamma_{\mathrm{safe}}}
\newcommand{\gammaunsafe}{\gamma_{\mathrm{unsafe}}}

\newcommand{\ZN}{Z_{\mathcal{N}}}
\newcommand{\N}{{\mathcal{N}}}

\newcommand{\Zdynamics}{Z_{\mathrm{dyn}}}
\newcommand{\gammadynamics}{\gamma_{\mathrm{dyn}}}

\newcommand{\calH}{\mathcal{H}}

\title{Learning Robust Output Control Barrier Functions from Safe Expert Demonstrations\thanks{This work is funded in part by NSF award CPS-2038873 and CAREER award ECCS-2045834.} }
\author[1]{Lars Lindemann}
\author[2]{Alexander Robey}
\author[3]{Lejun Jiang}
\author[2]{Satyajeet Das}
\author[2]{Stephen Tu}
\author[4]{Nikolai Matni}
\affil[1]{Thomas Lord Department of Computer Science, University of Southern California}
\affil[2]{Department of Electrical and Systems Engineering, University of Pennsylvania}
\affil[3]{Nuro, Inc.}
\affil[4]{Ming Hsieh Department of Electrical and Computer Engineering, University of Southern California}

\begin{document}

\maketitle

\begin{abstract}
This paper addresses learning safe output feedback control laws from partial observations of expert demonstrations. We assume that a model of the system dynamics and a state estimator are available along with corresponding error bounds, e.g., estimated from data in practice. We first propose  robust output control barrier functions (ROCBFs) as a means to guarantee safety, as defined through controlled forward invariance of a safe set. We then formulate an optimization problem to learn ROCBFs from expert demonstrations that exhibit safe system behavior, e.g., data collected from a human operator or an expert controller. When the parametrization of the ROCBF is linear, then we show that, under mild assumptions, the optimization problem is convex. Along with the optimization problem, we provide verifiable conditions in terms of the density of the data, smoothness of the system model and state estimator, and the size of the error bounds that guarantee validity of the obtained ROCBF. Towards obtaining a practical control algorithm, we  propose an algorithmic implementation of our theoretical framework that accounts for assumptions made in our framework in practice.  We  validate our algorithm in the autonomous driving simulator CARLA and demonstrate how to learn safe control laws from simulated RGB camera images.
\end{abstract}


\section{Introduction}

Safety-critical systems rely on robust control laws that can account for uncertainties in system dynamics and state estimation. For example, consider an autonomous car equipped with noisy sensors that navigates through urban traffic~\cite{schwarting2018planning}. The state of the car is not exactly known and  estimated from output measurements, e.g., from a dashboard camera, while the dynamics of the car are not perfectly known either, e.g., due to unknown friction coefficients. A model of the system dynamics and a state estimator can usually be obtained, e.g., from first principles or estimated from data, along with uncertainty sets describing error bounds. Such error bounds are standard in robust control theory \cite{freeman2008robust}, but designing robust control for described systems is  challenging. In this paper, we address this problem by using the increasing  availability of safe expert demonstrations, e.g., car manufacturers recording safe driving behavior of expert drivers. We  propose  a data-driven approach  to  learning safe and robust control laws where safety is defined as the ability of a system to stay within a set of states that are labeled safe, e.g., states that satisfy a minimum safety distance.

\subsection{Related Work}
\label{sec:related_work}

Control barrier functions (CBFs) were introduced in \cite{wieland2007constructive,ames2017control} to render  a safe set controlled forward invariant. A CBF defines a set of safe control inputs that can be used to find a minimally invasive safety-preserving correction to a nominal control law by solving a convex quadratic program. Many variations and extensions of CBFs  appeared in the literature, e.g., composition of CBFs \cite{glotfelter2017nonsmooth}, CBFs for multi-robot systems \cite{wang2017safety}, CBFs encoding temporal logic constraints \cite{lindemann2018control}, and CBFs for systems with higher relative degree \cite{xiao2019control}. Finally,  CBFs and Hamilton-Jacobi were found to share connections ~\cite{choi2021robust}.

CBFs that account for uncertainties in the system dynamics have been considered in two ways. The authors in \cite{kolathaya2018input} and \cite{xu2015robustness} consider input-to-state safety to quantify possible safety violation. Conversely, the work in \cite{jankovic2018robust} proposes robust CBFs  to guarantee robust safety by accounting for all permissible errors within an uncertainty set. Input delays within CBFs were discussed in \cite{seiler2021control,quan2023tube}.  CBFs that account for state estimation uncertainties were  proposed in \cite{dean2020guaranteeing} and \cite{zhang2021control}. Relying on the same notion of measurement robust CBFs as in \cite{dean2020guaranteeing}, the authors in \cite{cosner2021measurement} present empirical evaluations on a segway. While the notion of ROCBFs  that we present in this paper is inspired by measurement-robust CBFs as presented in \cite{dean2020guaranteeing}, we also consider uncertainties in the system dynamics and focus on learning valid CBFs from expert demonstrations. Similar to the notion of ROCBF, the authors in \cite{garg2021robust} consider additive disturbances in the system dynamics and state-estimation errors jointly. 

\textbf{Learning with CBFs:} Approaches that use CBFs during learning typically assume that a valid CBF is already given, while we focus on constructing CBFs so that our approach can be viewed as complementary.   In \cite{massiani2021safe}, it is shown how safe and optimal reward functions can be obtained, and how these are related to CBFs. The authors in \cite{ferlez2020shieldnn} use CBFs to learn a provably correct neural network safety guard for kinematic bicycle models. The authors in \cite{lopez2020robust} consider that uncertainty enters the system dynamics linearly and propose to use robust adaptive CBFs, as originally presented in \cite{taylor2020adaptive}, in conjunction with online set membership identification methods. In \cite{emam2021data}, it is shown how additive and multiplicative noise can be estimated online using Gaussian process regression for safe CBFs. The authors in \cite{taylor2020learning}  collect data to episodically update the system model and  the CBF controller. A similar idea is followed in \cite{csomay2021episodic} where instead a projection with respect to the CBF condition is episodically learned. Imitation learning under safety constraints imposed by a Lyapunov function was proposed in \cite{yin2021imitation}. Further work in this direction can be found in
\cite{cheng2019end,wang2018safe,verginis21}. 

\textbf{Learning CBFs: } An open problem is how valid CBFs can be constructed. Indeed, the lack of systematic methods to construct valid CBFs is a main bottleneck.  For certain types of mechanical systems under input constraints, analytic CBFs can be constructed \cite{cortez2020correct}.  The construction of polynomial barrier functions towards certifying safety for polynomial systems by using sum-of-squares (SOS) programming was proposed in \cite{prajna2007framework}. Finding CBFs poses additional challenges in terms of the control input resulting in bilinear SOS programming as presented in \cite{xu2017correctness,wang2018permissive} and summarized in \cite{ames2019control}. The work in \cite{clark2021verification} considers the construction of higher order CBFs and their composition by, similarly to \cite{xu2017correctness,wang2018permissive},  alternating-descent heuristics to solve the arising bilinear SOS program. Such SOS-based approaches, however, are known to be limited in scalability and do not use potentially available expert demonstrations.

A promising research direction is to learn CBFs from data. The authors in \cite{srinivasan2020synthesis} construct CBFs from safe and unsafe data using support vector machines, while authors in \cite{saveriano2019learning} learn a set of linear CBFs for clustered datasets. The authors in \cite{ohnishi2021constraint} proposed learning limited duration CBFs and the work in \cite{long2021learning} learns signed distance fields that define a CBF. In \cite{yaghoubi2020training}, a neural network controller is trained episodically to imitate an already given CBF. The authors in \cite{xiao2020feasibility} learn parameters associated with the constraints of a CBF to improve feasibility. These works present empirical validations, but no formal correctness guarantees are provided.  The authors in \cite{chen2020learning,abate2021fossil,dai2020counter,kapinski2014simulation} propose counter-example guided approaches to learn Lyapunov and barrier functions for known closed-loop systems, while Lyapunov functions for unknown systems are learned in \cite{boffi2020learning}. In~\cite{jin2020neural,dawson2022safe,zhao2020learning} control barrier functions are learned and post-hoc verified, e.g., using Lipschitz arguments and satisfiability modulo theory, while \cite{chang2019neural} uses a counter-example guided approach. As opposed to these works, we make use of safe expert demonstrations. Expert trajectories are utilized in  \cite{sun2020learning} to learn a contraction metric along with a tracking controller, while  motion primitives are learned from expert demonstrations in \cite{khansari2014learning}. In our previous work \cite{robey2020learning}, we proposed to learn CBFs for known nonlinear systems from expert demonstrations. We provided the first conditions that ensure correctness of the learned CBF using Lipschitz continuity and covering number arguments. In \cite{robey2021learning} and \cite{lindemann2020learning}, we extended this framework to partially unknown hybrid systems. In this paper, we focus on state estimation and provide sophisticated simulations of our method in CARLA. 

\subsection{Contributions}

In this paper, we learn safe output feedback control laws for unknown systems. We first present robust output control barrier functions (ROCBFs) to establish safety under system dynamics and state estimation uncertainties. We then formulate a constrained optimization problem for constructing ROCBFs from safe expert demonstrations, and we present verifiable conditions  that  guarantee the validity of the  ROCBF. While the optimization problem is in general nonconvex, we identify conditions under which the problem is convex. For the general case, we propose an approximate unconstrained  optimization problem that we can solve efficiently. Finally, we propose an algorithmic implementation of our theoretical framework to learn ROCBFs in practice, and we  present an empirical validation in CARLA \cite{dosovitskiy2017carla}.

In contrast to our previous works \cite{robey2020learning,lindemann2020learning,robey2021learning}, in which we assume perfect state knowledge, we focus on dealing with state estimation errors. Our paper additionally differs from \cite{robey2020learning,lindemann2020learning,robey2021learning} in its practical focus. We discuss the algorithmic implementation of our framework to account for assumptions of our work in practice. For instance, our framework crucially relies on obtaining ``unsafe'' data which is hard to obtain in practice, and we propose a new algorithm to obtain unsafe datapoints as boundary points from the set of safe expert demonstrations based on reverse $k$-nearest neighbors. 


\section{Background and Problem Formulation}
\label{sec:problem}

At time $t\in\RR_{\ge 0}$, let $\xx(t)\in\RR^\statedim$  be the state of the dynamical control system described by the set of equations
\begin{subequations}\label{eq:system}
	\begin{align}
		\dot{\xx}(t)&=\f(t)+\g(t)\uu(t), \\
		\xx(0)&:=x_0 \,\;\;\;\;\;\;\;\;\;\;\;\;\;\;\;\;\;\,\text{(initial condition),}\\
		f(t)&:=F(x(t),t)\,\;\;\;\;\;\;\,\text{(internal dynamics),}\label{eq:f}\\
		g(t)&:=G(x(t),t)\,\;\;\;\;\;\;\,\text{(input dynamics),}\\
		y(t)&:=Y(x(t)) \;\;\;\;\;\;\;\;\;\;\text{(output measurements),} \label{eq:y}\\
		u(t)&:=U(y(t),t) \;\;\;\;\;\;\,\, \text{(output feedback control law),} \label{eq:u}
	\end{align}
\end{subequations}
where $x_0\in \RR^\statedim$  is the 
initial condition. The functions $F:\RR^\statedim\times\mathbb{R}_{\ge 0}\to\RR^\statedim$ and $G:\RR^\statedim\times\mathbb{R}_{\ge 0}\to\RR^{n\times\inputdim}$ are  \emph{only partially known}, e.g., due to unmodeled dynamics or noise, and locally Lipschitz continuous in the first and piecewise continuous and bounded in the second argument. 
\vspace{-0.25cm}
\begin{assumption}\label{ass:1}We assume known nominal models $\hat{F}:\mathbb{R}^n\times\mathbb{R}_{\ge 0}\to\mathbb{R}^{n}$ and $\hat{G}:\mathbb{R}^n\times\mathbb{R}_{\ge 0}\to\mathbb{R}^{n\times m}$  together with functions $\Delta_F:\mathbb{R}^n\times\mathbb{R}_{\ge 0}\to\mathbb{R}_{\ge 0}$ and $\Delta_G:\mathbb{R}^n\times\mathbb{R}_{\ge 0}\to\mathbb{R}_{\ge 0}$ that bound their respective errors as\footnote{We let $\|\cdot\|$ be a vector norm and denote $\|\cdot\|_\star$ by its dual norm, while $\vertiii{\cdot}$ is the induced matrix norm.}
	\begin{align*}
		\|\hat{F}(x,t)-F(x,t)\|&\le \Delta_F(x,t) \text{ for all } (x,t)\in \mathbb{R}^n\times \mathbb{R}_{\ge 0},\\
		\vertiii{\hat{G}(x,t)-G(x,t)}&\le \Delta_G(x,t) \text{ for all } (x,t)\in \mathbb{R}^n\times \mathbb{R}_{\ge 0}.
	\end{align*} 
	The functions $\hat{F}(x,t)$, $\hat{G}(x,t)$, $\Delta_F(x,t)$ and $\Delta_G(x,t)$ are assumed to be locally Lipschitz continuous in the first and piecewise continuous and bounded in the second argument.
\end{assumption}
\vspace{-0.25cm}
The models $\hat{F}(x,t)$ and $\hat{G}(x,t)$ may be obtained  by identifying model parameters or by system identification \cite{ljung1998system}, while the assumption of error bounds $\Delta_F(x,t)$ and $\Delta_G(x,t)$ is standard in robust control \cite{freeman2008robust}. We now define the set of \emph{admissible system dynamics} as
\begin{align*}
	\mathcal{F}(x,t)&:=\{f\in\mathbb{R}^{n}|\|\hat{F}(x,t)-f\|\le \Delta_F(x,t)\},\\
	\mathcal{G}(x,t)&:=\{g\in\mathbb{R}^{n\times m}|\vertiii{\hat{G}(x,t)-g}\le \Delta_G(x,t)\}.
\end{align*}

\begin{figure}[H]
	\centering
	\includegraphics[scale=0.25]{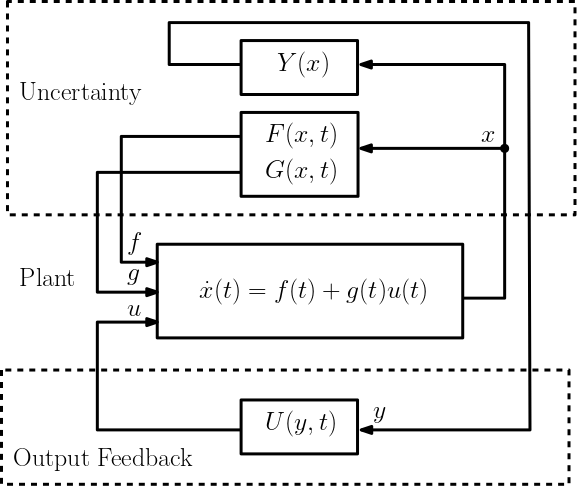}
	\caption{Uncertain system under consideration.}
	\label{fig:overview}
\end{figure}

The output measurement map $Y:\mathbb{R}^n\to\mathbb{R}^p$  is  \emph{only partially known} and locally Lipschitz continuous. For instance, $Y$ can describe a dashboard camera that is  hard to model. We assume that there exists an inverse yet unknown map $X:\mathbb{R}^p\to\mathbb{R}^n$ that recovers the state $x$ as $X(Y(x))=x$. This  means that a measurement $y$ uniquely defines a corresponding state $x$ and  implies that $p\ge n$. This way, we implicitly assume  high-dimensional measurements $y$ such as a dashboard camera where the inverse map $X$ recovers the position of the system, or even its velocity when a sequence of camera images is available. Since $Y$ and $X$ are unknown, one can, however, not recover the state $x$ from $y$.  We  present an example in the simulation study and  refer to related literature using similar assumptions,  such as \cite{dean2020guaranteeing,dean2020robust}.  
\vspace{-0.25cm}
\begin{assumption}\label{ass:2}We assume to have a known  model $\hat{X}:\mathbb{R}^p \to\mathbb{R}^n$  together with a function $\Delta_X:\mathbb{R}^p\to\mathbb{R}_{\ge 0}$ that bounds the error 
	\begin{align*}
		\|\hat{X}(y)-X(y)\|&\le \Delta_X(y) \text{ for all } y\in Y(\mathbb{R}^n).
	\end{align*} 
	The  functions $\hat{X}(y)$ and $\Delta_X(y)$ are assumed to be locally Lipschitz continuous.
\end{assumption}
\vspace{-0.25cm}
The state estimator $\hat{X}(y)$ and the error bound $\Delta_X(y)$ may  be obtained using machine learning methods, see e.g., \cite{dean2020guaranteeing,dean2020robust}, or $\hat{X}(y)$ can encode the extended Kalman filter together with $\Delta_X(y)$, see e.g., \cite{chou2022safe,kohler2021robust}. We now define the set of \emph{admissible inverse output measurement maps} as
\begin{align*}
	\mathcal{X}(y)&:=\{x\in\mathbb{R}^{n}|\|\hat{X}(y)-x\|\le \Delta_X(y)\}.
\end{align*}
Finally, the function $U:\mathbb{R}^p\times\mathbb{R}_{\ge 0}\to \mathcal{U}$ is the output feedback control law where $\mathcal{U}\subseteq\mathbb{R}^m$ encodes input constraints. System \eqref{eq:system} is illustrated in Fig. \ref{fig:overview}. Let a solution to \eqref{eq:system}  under an output feedback control law $U(y,t)$ be $\xx:\I\to \RR^\statedim$ where $\I\subseteq\RR_{\ge 0}$ is the maximum definition interval of $\xx$. 
\vspace{-0.5cm}

The goal in this paper is to learn an output feedback control law $U(y,t)$ such that prescribed safety properties with respect to a \emph{geometric safe set} $\mathcal{S}\subseteq \RR^\statedim$ are met by the system in \eqref{eq:system}. By geometric safe set, we mean that $\mathcal{S}$  describes the set of safe states as naturally specified on a subset of the state space (e.g., to avoid collision, vehicles must maintain a minimum separating distance). 
\vspace{-0.25cm}
\begin{definition}
	A set $\mathcal{C}\subseteq\mathbb{R}^{n}$ is said to be \emph{robustly output controlled forward invariant} with respect to the system in \eqref{eq:system} if there exist an output feedback control law $U(y,t)$ such that, for all initial conditions $x(0)\in\mathcal{C}$, for all admissible system dynamics $F(x,t)\in \mathcal{F}(x,t)$ and $G(x,t)\in \mathcal{G}(x,t)$, and for all  admissible inverse output measurement maps $X(y)\in\mathcal{X}(y)$, every solution $x(t)$ to \eqref{eq:system} under  $U(y,t)$
	is such that: 1) $x(t)\in\mathcal{C}$ for all $(t)\in \mathcal{I}$, and 2) the interval $\mathcal{I}$ is unbounded, i.e., $\I=[0,\infty)$. If the set $\mathcal{C}$ is additionally contained within the geometric safe set $\mathcal{S}$, i.e., $\mathcal{C}\subseteq\mathcal{S}$, the system in \eqref{eq:system} is said to be \emph{safe} under the \emph{safe control law} $U(y,t)$.
\end{definition}
\vspace{-0.25cm}
Towards this goal, we assume a data set of  \emph{expert demonstrations} consisting of $N_1$ input-output data pairs $(y_i,u_i)\in\mathbb{R}^p\times\mathbb{R}^m$ along with a time stamp $t_i\in\mathbb{R}_{\ge 0}$~as
\begin{align*}
	\Zdynamics:=\{(y_i,t_i,u_i)\}_{i=1}^{N_1}
\end{align*}
that were recorded when the system was in a safe state $X(y_i)\in\text{int}(\mathcal{S})$ where $\text{int}(\mathcal{S})$ denotes the interior of the safe set $\mathcal{S}$. We assume to have  expert control inputs $\uu_i$ available that can later be used for learning a safe control law. The pairs of expert demonstrations $(y_i,t_i,u_i)$ have to be such that a system trajectory starting from a state $x\in \Delta_X(y_i)$ can be kept within the safe set $\mathcal{S}$. If this was not the case, the later posed optimization problem (in equation \eqref{eq:opt}) would be infeasible. There are interesting observations as to  what constitutes a ``good'' expert action $u_i$, see  \cite{robey2020learning} for details.

\begin{figure}[H]
	\centering
	\includegraphics[scale=0.2]{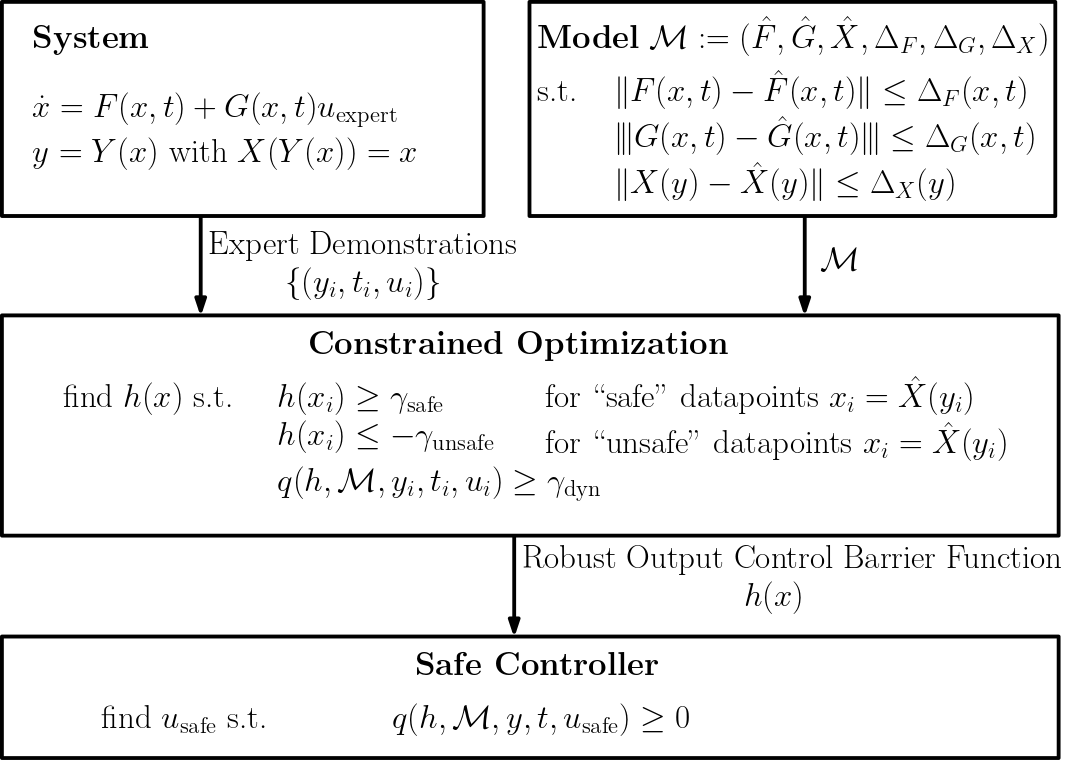}
	\caption{Proposed framework to learn safe control laws. }
	\label{fig:overview_}
\end{figure}

\vspace{-0.25cm}
\begin{problem}\label{prob:1}Let the system in \eqref{eq:system} and  the set of safe expert demonstrations $\Zdynamics$ be given. Under Assumptions \ref{ass:1} and \ref{ass:2}, learn a function $h:\mathbb{R}^n\to\mathbb{R}$ from $\Zdynamics$ so that the set
	\begin{align} \label{eq:set_C}
		\C:=\{\xx\in\mathbb{R}^n\, \big{|} \, \cbf(\xx)\ge 0\}
	\end{align}
	is robustly output controlled forward invariant  with respect to \eqref{eq:system} and such that $\mathcal{C}\subseteq\mathcal{S}$, i.e., so that  \eqref{eq:system} is safe. 
\end{problem}
\vspace{-0.25cm}
An overview of our proposed solution  is shown in Fig.~\ref{fig:overview_}. We formulate a constrained optimization problem to learn a function $h(x)$ so that the learned safe set $\mathcal{C}$ is robustly output controlled forward invariant and  contained within the geometric safe set $\mathcal{S}$, i.e., $\mathcal{C}\subseteq\mathcal{S}$. The optimization problem takes the system model $\mathcal{M}:=(\hat{F},\hat{G},\hat{X},\Delta_F,\Delta_G,\Delta_X)$ and the expert demonstrations $\{(y_i,u_i,t_i)\}$ as inputs and imposes constraints on a function $q$ and $h(x)$ that will be derived in the sequel.  We remark that the proofs of technical lemmas, propositions, and theorems  can be found in the appendices.

\section{Robust Output Control Barrier Functions}
\label{sec:rocbf}
Let $\cbf:\RR^\statedim\to\RR$ be a twice continuously differentiable function, and assume that $\cbf$ is such that the set $\C$ in \eqref{eq:set_C} has non-empty interior. Let $\mathcal{Y}\subseteq\mathbb{R}^p$ be a sufficiently large open set such that $\mathcal{Y}\supseteq Y(\C)$ where $Y(\C)$ denotes the image of $\mathcal{\C}$ under $Y$.\footnote{Note that the function $Y$ is unknown. In Section \ref{sec:optim}, we construct the sets $\mathcal{Y}$ and $\mathcal{C}$  in a way so that $\mathcal{Y}\supseteq Y(\C)$ holds.} This assumption is equivalent to $X(\mathcal{Y})\supseteq C$. The set $X(\mathcal{Y})$  is typically the domain of interest in existing state-based CBF frameworks, see e.g., \cite{ames2017control}. 

We recall that a function  $\alpha:\RR\to\RR$ is an extended class $\mathcal{K}$ function if it is a strictly increasing function with $\alpha(0)=0$. We can guarantee that $\mathcal{C}$ is robustly output controlled forward invariant if there exists a locally Lipschitz continuous extended class $\mathcal{K}$ function $\alpha:\RR\to\RR$ such that
\begin{align}\label{eq:cbf_const}
	\begin{split}
		\sup_{\uu\in \U} \inf_{x\in \mathcal{X}(y)} \inf_{\substack{f\in\mathcal{F}(x,t)\\g\in\mathcal{G}(x,t)}}  
		&\underbrace{\langle \nabla \cbf(\xx), f+gu\rangle}_{\substack{\text{Change in } h \text{ along} \\\text{all dynamics}}}+\alpha(\cbf(\xx)) \ge 0
	\end{split}
\end{align} 
for all $(y,t)\in\mathcal{Y}\times\mathbb{R}_{\ge 0}$ where $\langle \cdot,\cdot\rangle$ denotes the inner-product between two vectors. Unfortunately, the condition \eqref{eq:cbf_const} is difficult to evaluate due  the infimum operators. Towards a more tractable condition, we first define the function $B:\mathbb{R}^n\times\mathbb{R}_{\ge 0}\times\mathcal{U}\to\mathbb{R}$ as
\begin{align*}
	B(x,t,u)&:=\underbrace{\langle\nabla \cbf(\xx), \hat{F}(x,t)+\hat{G}(x,t)u\rangle}_{\text{Change in }h \text{ along model dynamics}}+\alpha(\cbf(\xx))\underbrace{-\|\nabla \cbf(\xx)\|_\star(\Delta_F(x,t)+\Delta_G(x,t)\|u\|)}_{\text{Robustness term accounting for system model error} }
\end{align*}
where $\|\cdot\|_\star$ denotes the dual norm. Satisfaction of constraint  \eqref{eq:cbf_const} can now  be guaranteed by
\begin{align*}
	\sup_{\uu\in \U} \inf_{x\in \mathcal{X}(y)} B(x,t,u) \ge 0
\end{align*}
reducing the complexity to the  infimum operator over the state measurement uncertainty $\mathcal{X}(y)$. For an output measurement $y\in\mathcal{Y}$ and fixed $u$ and $t$, denote  the local Lipschitz constant of the function $B(x,t,u)$ within the set  $\mathcal{X}(y)\subseteq\mathbb{R}^n$ by $\text{Lip}_B(y,t,u)$. We now define ROCBF that will guarantee that the set $\mathcal{C}$ is robustly output controlled forward invariant.



\vspace{-0.25cm}
\begin{definition}\label{def:rocbf}The function $\cbf(\xx)$ is said to be a \emph{robust output control barrier function} (ROCBF) on an open set $\mathcal{Y}\supseteq Y(\mathcal{C})$ if there exist a locally Lipschitz continuous extended class $\mathcal{K}$ function $\alpha:\RR\to\RR$\footnote{Recall that $\alpha$ is contained within the function $B(x,t,u)$.} such that 
	\begin{align}\label{eq:cbf_const_}
		\begin{split}
			\sup_{\uu\in \U} &\; B(\hat{X}(y),t,u) -\underbrace{\text{Lip}_B(y,t,u)\Delta_X(y)}_{\substack{\text{Robustness term accounting}\\ \text{for state estimation error} }}\ge0
		\end{split}
	\end{align} 
	for all $(y,t)\in\mathcal{Y}\times\mathbb{R}_{\ge 0}$.
\end{definition}
\vspace{-0.25cm}
Note that ROCBFs account for both system model and estimation error uncertainties. The standard CBF condition from \cite{ames2017control} is recovered if the system  is completely known, i.e., the sets $\mathcal{F}(x,t)$, $\mathcal{G}(x,t)$, and $\mathcal{X}(y)$ are singletons.  Now define the set of safe control inputs induced by a ROCBF as
\begin{align*}
	\mathcal{U}_\text{s}(y,t)&:=\{\uu\in\mathcal{U} \, \big{|} \, 
	B(\hat{X}(y),t,u)-\text{Lip}_B(y,t,u)\Delta_X(y)\ge0 \}.
\end{align*}



We next show that a control law $U(y,t)\in \mathcal{U}_\text{s}(y,t)$ renders the set $\C$ robustly output controlled forward invariant. 
\vspace{-0.25cm}
\begin{theorem}\label{them:1}Assume that $\cbf(x)$ is a ROCBF on the set $\mathcal{Y}$ that is such that $\mathcal{Y}\supseteq Y(\mathcal{C})$, and assume that the function $U:\mathcal{Y}\times \mathbb{R}_{\ge 0}\to \U$ is continuous in the first and piecewise continuous in the second argument and such that $U(y,t)\in \mathcal{U}_\text{s}(y,t)$. Then $\xx(0)\in\C$ implies $\xx(t)\in\C$ for all $t\in \I$. If the set $\C$ is compact, it follows that $\C$ is robustly output controlled forward invariant under $U(y,t)$, i.e., $\I=[0,\infty)$.
\end{theorem}
\vspace{-0.25cm}





\section{Learning ROCBFs from Expert Demonstrations}
\label{sec:optim}

The previous section provides safety guarantees when $h(x)$ is a ROCBF. However, one is still left with the potentially difficult task of constructing a twice continuously differentiable function $h(x)$ such that (i) the set $\C$ defined in equation \eqref{eq:set_C} is contained within the set $\mathcal{S}$ and has a sufficiently large volume, and (ii) it satisfies the barrier constraint \eqref{eq:cbf_const_} on an open set $\mathcal{Y}$ that is such that $\mathcal{Y}\supseteq Y(\C)$.  In fact, ensuring that a function $h(x)$ satisfies the constraint \eqref{eq:cbf_const_} can involve verifying complex relationships between the vector fields $\hat{F}(x,t)$ and $\hat{G}(x,t)$, the state estimate $\hat{X}(y)$,  the function $h(x)$, and its gradient $\nabla h(x)$, while accounting for the error bounds $\Delta_X(y)$ as well as $\Delta_F(x,t)$ and $\Delta_G(x,t)$. 

This challenge motivates the approach taken in this paper, wherein we propose an optimization-based approach to learning a ROCBF from safe expert demonstrations.


\subsection{The Datasets }
\label{sec:datasets}

\begin{figure*}
	\centering
	\includegraphics[scale=0.1375]{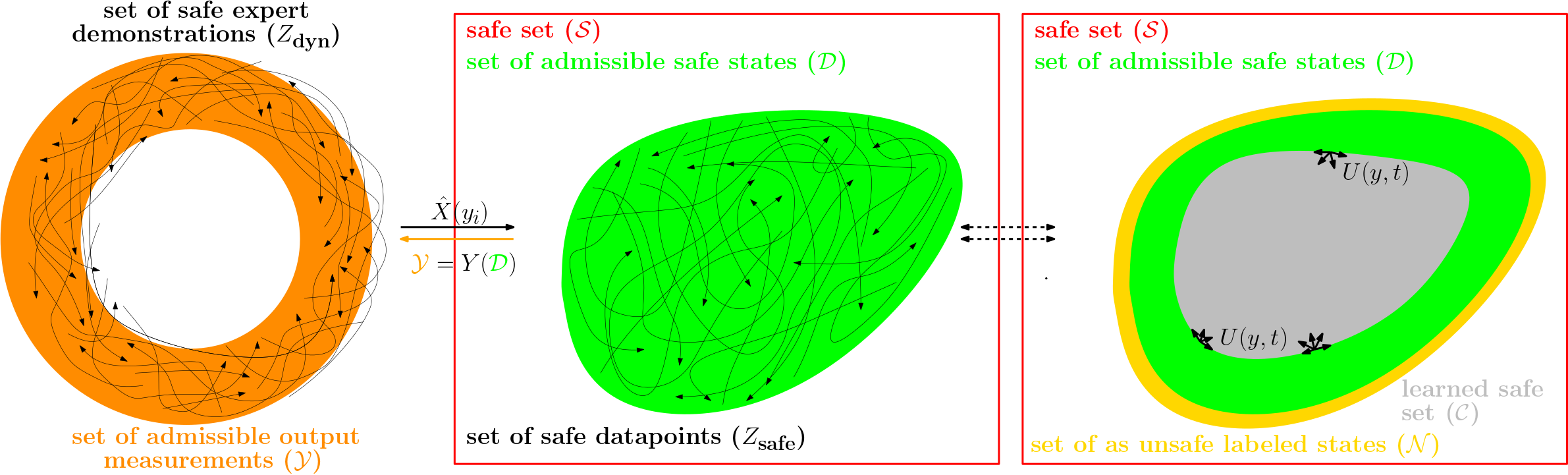}
	\caption{Problem Setup (left): The set of observed safe expert demonstrations $\Zdynamics$ (black lines). Also shown, the set of admissible output measurements $\mathcal{Y}$ (orange ring).  Transformation into state domain (centre): The geometric safe set $\Ss$ (red box) and the set of admissible safe states $\mathcal{D}$ (green region) that is defined as the union of $\epsilon$-balls centered at $\hat{X}(y_i)$. Learned safe set (right): The set of as unsafe labeled states $\N$ (golden ring) that defines the $\sigma$-layer surrounding $\D$.}
	\label{fig:1}
\end{figure*}

We first define the finite set of safe datapoints
\begin{align*}
	\Zsafe := \bigcup_{(y_i,t_i,u_i)\in \Zdynamics} \hat{X}(y_i)
\end{align*}
as the projection of all datapoints $y_i$ in $\Zdynamics$ via the state estimator $\hat{X}$ into the state domain.  For $\epsilon>0$, define the set of admissible states $\mathcal{D}\subseteq\mathbb{R}^n$ as
\begin{align*}
	\mathcal{D}&:=\mathcal{D}'\setminus \text{bd}(\mathcal{D}') \;\;\text{ with }\;\; \mathcal{D}':=\bigcup_{x_i\in \Zsafe}\B_{\epsilon}(x_i)
\end{align*}  
where $\B_{\epsilon}(x_i):=\{\xx\in\RR^\statedim\, \big{|}\, \|\xx-x_i\|\le \epsilon\}$ is the closed norm ball of size $\epsilon$ centered at  $x_i$ and where $\text{bd}(\cdot)$ denotes the boundary of a set. Conditions on $\epsilon$ will be specified later to ensure validity of the learned control law. The set $\mathcal{D}'$ is the union of these $\epsilon$ norm balls, see Fig. \ref{fig:1} (left and centre). The set of admissible states $\mathcal{D}$ is equivalent to the set $\mathcal{D}'$ without its boundary so that  $\mathcal{D}$ is open. Note that $\mathcal{D}$ is based on expert demonstrations $y_i$ via the state estimator $\hat{X}(y_i)$. The expert demonstrations $y_i$ in $\Zdynamics$ define an $\epsilon$-net of $\mathcal{D}$. In other words, for each $x \in \mathcal{D}$ there exists a $y_i$ in $(y_i,t_i,u_i)\in\Zdynamics$ such that $\|\hat{X}(y_i)-x\|\leq \epsilon$.  We additionally assume that  $\mathcal{D}$ is such that $\mathcal{D}\subseteq \mathcal{S}$, which can be easily achieved  by adjusting $\epsilon$ or by omitting $y_i$ from $\Zdynamics$ in the definition of $\Zsafe$ when datapoints $\hat{X}(y_i)$ are close to $\text{bd}(\mathcal{S})$. Note here that $\mathcal{S}$ is typically known as part of the safety specification. This additional requirement is necessary to later ensure safety in the sense that the learned safe set is such that $\mathcal{C}\subseteq\mathcal{S}$. 


We define the set of admissible output measurements as
\begin{align*}
	\mathcal{Y}:=Y(\mathcal{D}),
\end{align*}
i.e., as the projection of the set $\mathcal{D}$ under the unknown output measurement map $Y$. We remark that the set $\mathcal{Y}$, illustrated in Fig.~\ref{fig:1} (left),  is consequently also unknown.  Note however that the set $ \mathcal{Y}$ is open as required in Theorem \ref{them:1}.



For $\sigma>0$, we define the set of  unsafe labeled states 
\begin{align*}
	\N:=\{\text{bd}(\D)\oplus\B_{\sigma}(0)\} \setminus \D,
\end{align*} 
where $\oplus$ is the Minkowski sum operator. The set $\N$ should be thought of as a layer of width $\sigma$ surrounding the set $\D$, see Fig.~\ref{fig:1} (right) for a graphical depiction. As will be made clear in the sequel, by enforcing that the value of the learned function $\cbf(\xx)$ is negative on $\N$, we ensure that the set $\mathcal{C}$ (defined as the zero-superlevel set of $h(x)$) is contained within $\D$, and hence also within $\mathcal{S}$. This is why we refer to $\mathcal{N}$ as set of  unsafe labeled states. To ensure that $h(x)<0$ for all $x\in\mathcal{N}$, we assume that points 
\begin{align*}
	\ZN := \{\xx_i\}_{i=1}^{N_2}
\end{align*} 
are sampled from  $\N$ such that $\ZN$ forms an $\epsilon_\mathcal{N}$-net of $\N$, i.e., for each $x\in \N$ there exists a $x_i\in \ZN$ such that  $\|x-x_i\|\le \epsilon_\mathcal{N}$. Conditions on $\epsilon_\mathcal{N}$ will be specified later.   We emphasize that no  control inputs $\uu_i$ are needed for the samples in $\ZN $ as these points are not generated by the expert and are instead  obtained by  computational methods such as gridding or uniform sampling (see Section \ref{sec:compute_} for details).

While the definition of the set $\mathcal{C}$ in \eqref{eq:set_C} is specified over all of $\RR^n$, e.g., the definition of $\C$  considers all $x\in \RR^n$ such that $h(x) \geq 0$, we make a minor modification to this definition in order to restrict the domain of interest to $\N\cup\D$ as
\begin{align} \label{eq:local_C}
	\C:=\{\xx\in\N\cup\D\, \big{|} \, \cbf(\xx)\ge 0\}.
\end{align}

This restriction is natural, as we are learning a function $h(x)$ from data sampled only over $\N\cup\D$. The size of the set $\mathcal{D}$ affects the size of the set $\C$, i.e., $\C$ may be conservative if only few expert demonstrations are available, e.g., consider Fig.~\ref{fig:1} but with fewer expert demonstrations in which case the green and grey regions would simply shrink. 


\subsection{The Constrained Optimization Problem}

We first state  the constrained optimization problem for learning valid ROCBFs, and then provide conditions in Section~\ref{sec:conditions} under which a feasible solution is a valid ROCBF.

Let $\calH$ be a normed function space of twice continuously differentiable 
functions $h : \RR^n \to \RR$. Define
\begin{align}\label{eq_q_def}
	q(y,t,u)&:= B(\hat{X}(y),t,u) - \overline{\text{Lip}}_B(y,t,u)\Delta_X(y)
\end{align}
analogously to \eqref{eq:cbf_const_}, but using a known surrogate function $\overline{\text{Lip}}_B(y,t,u)$ in place of the Lipschitz constant ${\text{Lip}}_B(y,t,u)$. The function $\overline{\text{Lip}}_B(y,t,u)$ will be a hyperparameter\footnote{A natural choice is
	$\overline{\text{Lip}}_B(y,t,u):=\overline{\text{Lip}}_1+\overline{\text{Lip}}_2\|u\|_\star+\overline{\text{Lip}}_3\|u\|$
	for sufficiently large positive constants $\overline{\text{Lip}}_1$, $\overline{\text{Lip}}_2$, and $\overline{\text{Lip}}_3$.} in our algorithm as discussed in Section \ref{sec:compute_}, and will be adjusted to ensure that  $\overline{\text{Lip}}_B(y,t,u)\ge {\text{Lip}}_B(y,t,u)$.

We  formulate  the following constrained optimization problem to learn a ROCBF from expert demonstrations:
\begin{subequations}\label{eq:opt}
	\begin{align}
		&\argmin_{h \in \calH} \:\: \norm{h}   \:  \\
		&\mathrm{s.t.}~~ h(x_i) \geq \gammasafe, \hspace{0.825cm}\forall x_i \in \underline{Z}_\text{safe} \label{eq:opt_1}\\
		&~~~~~~ h(x_i) \leq -\gammaunsafe, \hspace{0.22cm} \forall x_i\in \ZN \label{eq:opt_2} \\
		&~~~~~~  q(y_i,t_i,u_i) \geq \gammadynamics, \forall (y_i,t_i,u_i) \in \Zdynamics \label{eq:opt_3}
	\end{align}
\end{subequations}
where the set $\underline{Z}_\text{safe}$ is a subset of $Z_\text{safe}$, i.e., $\underline{Z}_\text{safe}\subseteq Z_\text{safe}$, as detailed in the next section  and where  $\gammasafe, \gammaunsafe, \gammadynamics>0$ are
hyperparameters. Instead of global hyperparameters $\gammasafe$, $\gammaunsafe$, and $\gammadynamics$, one can use individual hyperparameters for each datapoint. Note that expert demonstrations $(y_i,t_i,u_i)$ indicate feasibility of the control problem at hand, and hence indicate feasibility of \eqref{eq:opt}. With increasing sizes of the uncertainty sets $\mathcal{F}(x,t)$, $\mathcal{G}(x,t)$, and $\mathcal{X}(y)$, the optimization problem \eqref{eq:opt} may however become infeasible.

\subsection{Conditions guaranteeing learned safe ROCBFs }
\label{sec:conditions}
We now derive  conditions under which a feasible solution to the constrained optimization problem \eqref{eq:opt} is a safe ROCBF. 

\subsubsection{Guaranteeing $\C\subset\D\subseteq\Ss$}
\label{sec:guarantees}
We begin with establishing the requirement that $\C\subset\D\subseteq\Ss$. First note that constraint \eqref{eq:opt_1} ensures that the set $\C$, as defined in equation \eqref{eq:local_C}, has non-empty interior when $\underline{Z}_\text{safe}\neq \emptyset$. We next state conditions under which the constraint \eqref{eq:opt_2} ensures that the learned function $h$ from \eqref{eq:opt}  satisfies $h(x) < 0$ for all $ x \in \N$, which in turn ensures that $\C\subset\D\subseteq\Ss$.

\vspace{-0.25cm}
\begin{proposition}\label{thm:0}Let $\cbf(\xx)$ be Lipschitz continuous with local Lipschitz constant $\text{Lip}_h(x_i)$ within the set $\mathcal{B}_{\epsilon_\mathcal{N}}(x_i)$ for datapoints $x_i \in \ZN$. Let $\gammaunsafe>0$, $\ZN$ be an $\epsilon_\mathcal{N}$-net of $\N$, and let 
	\begin{align}\label{thm:0_eq}
		\epsilon_\mathcal{N}<\frac{\gammaunsafe}{\text{Lip}_h(x_i)}
	\end{align} 
	for all $x_i \in \ZN$. Then, the constraint \eqref{eq:opt_2} ensures that $\cbf(\xx)<0$ for all $x\in\N$.
\end{proposition}
\vspace{-0.25cm}
In summary, Proposition \ref{thm:0} says that a larger Lipschitz constant of the function $h$ requires a larger margin $\gamma_\text{unsafe}$ and/or a finer net of unsafe datapoints as indicated by $\epsilon_\mathcal{N}$.

We next discuss the choice of $\underline{Z}_\text{safe}$. Assume first that $\underline{Z}_\text{safe}=Z_\text{safe}$ in  constraint \eqref{eq:opt_1}. In this case, the constraints \eqref{eq:opt_1} and \eqref{eq:opt_2}, as well as the condition in \eqref{thm:0_eq} of Proposition \ref{thm:0}, may be conflicting, leading to infeasibility of the optimization problem \eqref{eq:opt}. This infeasibility arises from the fact that we are simultaneously asking for the value of $\cbf(\xx)$ to vary from $\gammasafe$ to $-\gammaunsafe$ over a short distance of at most $\epsilon+\epsilon_\mathcal{N}$ while having a small Lipschitz constant.  In particular, as posed, the constraints require that $|h(x_s)-h(x_u)|\geq \gammasafe+\gammaunsafe$ for $x_s \in \Zsafe$ and $x_u\in \ZN$ safe and unsafe samples, respectively, but the sampling requirements ($\Zsafe$ and $\ZN$ being $\epsilon$ and $\epsilon_\mathcal{N}$-nets of $\mathcal{D}$ and $\mathcal{N}$, respectively) imply that $\|x_s-x_u\| \leq \epsilon_\mathcal{N}+\epsilon$ for at least some pair $(x_s,x_u)$, which in turn implies that 
\begin{align*}
	\text{Lip}_h(x_u) \gtrsim \frac{|h(x_s)-h(x_u)|}{\|x_s-x_u\|} \ge \frac{\gammasafe + \gammaunsafe}{\epsilon_\mathcal{N}+\epsilon}.
\end{align*}
The local Lipschitz constant $\text{Lip}_h(x_u)$ may hence get too large if $\gammasafe$ and $\gammaunsafe$ are chosen to be too large, and we may exceed the required upper bound $\gammaunsafe/\epsilon_\mathcal{N}$ in equation \eqref{thm:0_eq}. We address this issue as follows: for fixed $\gammasafe$, $\gammaunsafe$, and desired Lipschitz constant $\Lc<\gammaunsafe/\epsilon_{\mathcal{N}}$, we define
\begin{align}\label{eq:Z_safe_new}
	\hspace{-0.2cm}\underline{Z}_\text{safe}:=\big\{x_i \in \Zsafe  \big{|}  \inf_{x\in\ZN}\|\xx-\xx_i\| \ge \frac{\gammasafe+\gammaunsafe }{\Lc} \big\},
\end{align} 
which corresponds to a subset of admissible safe states, i.e., $\underline{Z}_\text{safe}\subset Z_\text{safe}$.  Intuitively, this introduces a buffer region across which $\cbf(\xx)$ can vary in value from $\gammasafe$ to $-\gammaunsafe$ for the desired Lipschitz constant  $\Lc$.  Enforcing  \eqref{eq:opt_1} over  $\underline{Z}_\text{safe}$ allows for smoother functions $h(x)$ to be learned at the expense of a smaller invariant safe set $\mathcal{C}$.

Note finally that, if  $h(x)$ is such that $\mathcal{C}\subset \mathcal{D}$ (i.e., under the conditions in Proposition \ref{thm:0}), then $Y(\mathcal{C})\subseteq Y(\mathcal{D})=\mathcal{Y}$ by definition of $\mathcal{Y}$ so that  $\mathcal{Y}\supseteq Y(\mathcal{C})$ as required in Theorem \ref{them:1}.

\subsubsection{Increasing the volume of $\mathcal{C}$}
We next explain how to avoid learning a safe set $\mathcal{C}$ consisting of many disconnected sets, which would not be practical, and show simultaneously how to increase the volume of $\mathcal{C}$. Let 
\begin{align*}
	\underline{\mathcal{D}}:=\cup_{x_i\in\underline{Z}_\text{safe}} \mathcal{B}_\epsilon(x_i)
\end{align*} 
and note that $\underline{Z}_\text{safe}$ is an $\epsilon$-net of $\underline{\mathcal{D}}$ by definition.  We next show conditions under which $h(x)\geq 0$ for all $x\in\underline{\mathcal{D}}$. 
\vspace{-0.25cm}
\begin{proposition}\label{cor:1}Let $\cbf(\xx)$ be Lipschitz continuous with local Lipschitz constant $\text{Lip}_h(x_i)$ within the set $\mathcal{B}_{{\epsilon}}(x_i)$ for datapoints $x_i\in \underline{Z}_\text{safe}$. Let $\gamma_\text{safe}>0$ and let
	\begin{align}\label{eq:propp}
		{\epsilon} \leq \frac{\gamma_\text{safe}}{\text{Lip}_h(x_i)}
	\end{align}
	for all $x_i\in \underline{Z}_\text{safe}$. Then, the constraint \eqref{eq:opt_1} ensures that $h(x)\geq 0$ for all $x\in \underline{\mathcal{D}}$.
\end{proposition}
\vspace{-0.25cm}
The previous result can be used to guarantee that the set $\C$ defined in equation \eqref{eq:local_C} contains the set $\underline{\mathcal{D}}$, i.e., $\underline{\mathcal{D}} \subseteq \C$. Hence, the set $\underline{\mathcal{D}}$ can be seen as the minimum volume of the set $\C$ that we can guarantee. Note that, under the provided conditions, it holds that $\C$ is such that $\underline{\mathcal{D}}\subseteq\C\subset\mathcal{D}\subseteq\mathcal{S}$.

We note that the amount of data needed to satisfy conditions \eqref{thm:0_eq} and \eqref{eq:propp} in Propositions \ref{thm:0} and \ref{cor:1} grows exponentially with the dimension $n$, see e.g., \cite[Section 4.2]{vershynin2018high}.

\subsubsection{Guaranteeing that $h(x)$ is a ROCBF}
\label{sec:theory:valid_local}

Propositions \ref{thm:0} and \ref{cor:1} guarantee that the level-sets of the learned function $h(x)$ satisfy the desired geometric safety properties.  We now derive conditions that ensure that $h(x)$ is a ROCBF, i.e., that the ROCBF constraint \eqref{eq:cbf_const_} is also satisfied. 

To satisfy constraint \eqref{eq:cbf_const_} for each $(y,t) \in \mathcal{Y}\times \mathbb{R}_{\ge 0}$, there must exist a control input $u\in\mathcal{U}$ such that $q(y,t,u)\geq 0$. We follow a similar idea as in Propositions \ref{thm:0} and \ref{cor:1} and note in this respect that the $y$ components of $\Zdynamics$ form an $\bar{\epsilon}$-net of $\mathcal{Y}$ (see Appendix \ref{sec:F} for a proof) where $\bar{\epsilon}$ is 
\begin{align*}
	\bar{\epsilon}:=\text{Lip}_Y(\epsilon+\bar{\Delta}_X)
\end{align*} 
with $\bar{\Delta}_X:=\sup_{y\in \mathcal{Y}}\Delta_X(y)$ denoting the maximum  estimation error and $\text{Lip}_Y$ being the  Lipschitz constant of the output measurement map $Y$ within the set     $\overline{\mathcal{D}}:=\mathcal{D}\oplus \mathcal{B}_{2\bar{\Delta}_X}(0)$.\footnote{The  set $\overline{\mathcal{D}}$ is equivalent to the set of admissible safe states $\mathcal{D}$ enlarged by a ball of size $2\bar{\Delta}_X$.} 

We additionally assume to know a bound on the difference of the function $q$ for different times $t$. More formally, for each $\bar{y}\in \mathcal{B}_{\bar{\epsilon}}(y)$, let $\text{Bnd}_q(y,u)$ be such that 
\begin{align}\label{eq:bound_q}
	|q(\bar{y},t',u)-q(\bar{y},t'',u)|\le \text{Bnd}_q(y,u), \forall t',t''\ge 0.
\end{align}
The bound $\text{Bnd}_q(y,u)$ exists and can be obtained as all components of $q(y,t,u)$ are bounded in $t$. This is a natural assumption to obtain formal guarantees on the function $q(y,t,u)$ from a finite  dataset $Z_\text{dyn}$  since it is not possible to sample the time domain $\mathbb{R}_{\ge 0}$ densely with a finite number of samples. It can be seen that $\text{Bnd}_q(y,u)=0$ when the system \eqref{eq:system} is independent of~$t$.

\vspace{-0.25cm}
\begin{proposition}\label{prop:2}Let $q(y,t,u)$  be Lipschitz continuous\footnote{Note that $q(y,t_i,u_i)$ is locally Lipschitz continuous. As the function $\cbf(\xx)$ is twice continuously differentiable, we immediately have that $\nabla\cbf(\xx)$ is locally Lipschitz continuous over the bounded domain $\mathcal{D}$. Also note that $\hat{F}$, $\hat{G}$, $\Delta_F$, $\Delta_G$, $\alpha$, $h(x)$, $\hat{X}$, and $\Delta_X$ are Lipschitz continuous.}  in $y$ for fixed $t$ and $u$ with local Lipschitz constant $\text{Lip}_q(y_i,t_i,u_i)$ within the set $\mathcal{B}_{\bar{\epsilon}}(y_i)$ for each $(y_i,t_i,u_i)\in Z_\text{dyn}$. Let $\gamma_\text{dyn}>0$ and 
	\begin{align}\label{thmmmm}
		\bar{\epsilon}\leq \frac{\gamma_\text{dyn}-\text{Bnd}_q(y_i,u_i)}{\text{Lip}_q(y_i,t_i,u_i)} 
	\end{align}   
	for all $(y_i,t_i,u_i)\in Z_\text{dyn}$. Then, the constraint \eqref{eq:opt_3} ensures that,  for each $(y,t) \in \mathcal{Y}\times \mathbb{R}_{\ge 0}$, there exists a $u\in\mathcal{U}$ such that $q(y,t,u)\geq 0$. If additionally ${\text{Lip}}_B(y,t,u)\le \overline{\text{Lip}}_B(y,t,u)$ for each $(y,t,u) \in \mathcal{Y}\times \mathbb{R}_{\ge 0}\times \mathcal{U}$,  then $h(x)$ is a ROCBF.
\end{proposition}
\vspace{-0.25cm}
In summary, Proposition \ref{prop:2} says that a larger Lipschitz constant of the function $q$ requires a larger margin $\gamma_\text{dyn}$ and/or a smaller $\bar{\epsilon}$, i.e., a finer net of safe datapoints as indicated by $\epsilon$ and/or a reduction in the measurement map error $\Delta_X$.

The next theorem summarizes our results, follows from the previous results, and is provided without proof.  
\vspace{-0.25cm}
\begin{theorem}\label{thm:2}Let $h(x)$ be a twice continuously differentiable function. Let the sets $\mathcal{S}$, $\mathcal{C}$, $\mathcal{Y}$, $\underline{\mathcal{D}}$, and $\N$ as well as the data-sets $\underline{Z}_\text{safe}$, $\Zdynamics$, and $\ZN$ be defined as above. Suppose that $\ZN$ forms an $\epsilon$-net of $\N$ and that the conditions \eqref{thm:0_eq}, \eqref{eq:propp}, and \eqref{thmmmm} are satisfied. Assume also that  ${\text{Lip}}_B(y,t,u)\le \overline{\text{Lip}}_B(y,t,u)$ for each $(y,t,u) \in \mathcal{Y}\times \mathbb{R}_{\ge 0}\times \mathcal{U}$. If  $\cbf(\xx)$ satisfies the constraints \eqref{eq:opt_1}, \eqref{eq:opt_2}, and \eqref{eq:opt_3}, then $\cbf(\xx)$ is a ROCBF on $\mathcal{Y}$ and it holds that the set $\C$ is non-empty and such that $\underline{\mathcal{D}}\subseteq\C\subseteq\Ss$.
	\vspace{-0.25cm}
\end{theorem}

	\section{Algorithmic Implementation}
	\label{sec:compute_}
	
In this section, we present the algorithmic implementation of the previously presented results. We will discuss various aspects related to solving the constrained optimization problem \eqref{eq:opt}, the construction of the involved datasets, and estimating Lipschitz constants of the functions $h(x)$ and $q(y,t,u)$.

\subsection{The Algorithm}
\label{sec:algorithm}
We summarize our algorithm to learn safe ROCBFs $h(x)$ in Algorithm \ref{alg:overview}. We first construct the set of safe datapoints $Z_\text{safe}$  from the expert demonstrations $Z_\text{dyn}$ (line 3).  We construct the set of as unsafe labeled datapoints $Z_\mathcal{N}$ from  $Z_\text{safe}$ (line 4), i.e., $Z_\mathcal{N}\subseteq Z_\text{safe}$, by identifying boundary points in $Z_\text{safe}$ and labeling them as unsafe  (details can be found in Section \ref{sec:data}). We then re-define $Z_\text{safe}$ by removing the unsafe labeled datapoints $Z_\mathcal{N}$ from $Z_\text{safe}$ (line 5). Following our discussion in Section \ref{sec:conditions}, we obtain $\underline{Z}_\text{safe}$ according to equation \eqref{eq:Z_safe_new} (line 6). We then  solve the constrained optimization problem  \eqref{eq:opt} by an unconstrained relaxation defined in \eqref{eq:opt_relaxed} (line 7) as discussed in Section \ref{sec:compute}. Finally, we  check if the constraints \eqref{eq:opt_1}-\eqref{eq:opt_3} and the constraints \eqref{thm:0_eq}, \eqref{eq:propp}, \eqref{thmmmm} are satisfied by the learned function $h(x)$ (line 8).  If the constraints are not satisfied, the hyperparameters  are adjusted and the process is repeated (line 9). We discuss Lipschitz constant estimation of $h$ and $q$ and the hyperparameter selection in Section \ref{sec:Lipschitz_constants}.

\begin{algorithm}
	\centering
	\begin{algorithmic}[1]
		\Statex \textbf{Input: } Set of expert demonstrations $Z_\text{dyn}$,\\ system model $(\hat{F},\hat{G},\hat{X},\Delta_F,\Delta_G,\Delta_X)$,\\ hyperparameters $(\alpha,\gamma_\text{safe},\gamma_\text{unsafe},\gamma_\text{dyn},L_h,\overline{\text{Lip}}_{B},k,\eta)$
		\Statex \textbf{Output: } Safe ROCBF $h(x)$
		\State $Z_\text{safe} \gets \cup_{(y_i,t_i,u_i)\in Z_\text{dyn}}\hat{X}(y_i)$ \quad\# Safe datapoints
		\State $Z_\mathcal{N} \gets \textsc{BPD}(Z_\text{safe},k,\eta)$ \quad\# Unsafe datapoints obtained by boundary point detection (BPD) in Algorithm \ref{alg:mr-peanut}.
		\State $Z_\text{safe} \gets Z_\text{safe}\setminus Z_\mathcal{N}$ 
		\State $\underline{Z}_\text{safe} \gets $ according to \eqref{eq:Z_safe_new}
		\State $h(x) \gets $ solution of \eqref{eq:opt_relaxed}  \quad\#  relaxation of the constrained optimization problem in \eqref{eq:opt}
		\WHILE{constr. \eqref{eq:opt_1}-\eqref{eq:opt_3},  \eqref{thm:0_eq}, \eqref{eq:propp}, \eqref{thmmmm} are violated}
		\State Modify hyperparameters and start from line 3
		\ENDWHILE
	\end{algorithmic}
	\caption{Learning ROCBF from Expert Demonstrations}
	\label{alg:overview}
\end{algorithm}

While our algorithmic implementation is an approximate solution of the proposed framework, we mention that solving an unconstrained relaxation of \eqref{eq:opt} and bootstrapping hyperparameters is a common technique in machine learning when solving nonconvex constrained optimization problems \cite{chamon2020probably}. Such techniques are  necessary for learning based methods to be applied to realistic systems. As we  reported in earlier works, see e.g., \cite{robey2021learning} for hybrid systems, such techniques perform well in practice and can even outperform experts.

\subsection{Construction of the Datasets}
\label{sec:data}

Due to the conditions in equations \eqref{thm:0_eq}, \eqref{eq:propp}, and \eqref{thmmmm}, the first requirement  is that the datasets  $\underline{Z}_\text{safe}$ and $Z_\mathcal{N}$ are dense, i.e., that $\epsilon$ and $\epsilon_\mathcal{N}$ are small. It is also required that $Z_\mathcal{N}$ is an $\epsilon_\mathcal{N}$-net of the set of unsafe labeled states $\mathcal{N}$. In order to construct the $\epsilon_\mathcal{N}$-net $Z_\mathcal{N}$ of  $\mathcal{N}$,
a simple randomized algorithm, which repeatedly uniformly samples
from $\mathcal{N}$, works with high probability, see e.g., \cite{vershynin2018high}. Hence, as long as
we can efficiently sample from $\mathcal{N}$, e.g., when $\mathcal{N}$ is a basic primitive set or has a set-membership oracle, uniform sampling or gridding methods are viable strategies.  

However, as this is in general not possible, we  use a boundary point detection algorithm in line 4 of Algorithm \ref{alg:overview}. The idea is to obtain the set of unsafe labeled datapoints $Z_\mathcal{N}$ instead from the set of safe datapoints $Z_\text{safe}$.  To perform this step efficiently, our approach is to detect geometric boundary points of the set $\Zsafe$. This subset of boundary points is labeled as $Z_\mathcal{N}$, while we re-define $\Zsafe$ to exclude the boundary points $Z_\mathcal{N}$ in line 5 of Algorithm \ref{alg:overview}. Particularly, we detect boundary points in $\Zsafe$ based on the concept of reverse $k$-nearest neighbors, see e.g., \cite{xia2006border}. The main idea is that boundary points typically have fewer reverse $k$-nearest neighbors than interior points.  For $k>0$, we find the $k$-nearest neighbors of each datapoint $x_i\in\Zsafe$. Then, we find the reverse $k$-nearest neighbors of each datapoint $x_i\in\Zsafe$, that is, we find the datapoints $x_i'\in\Zsafe$ that have $x_i$ as their $k$-nearest neighbor. Finally, we choose a threshold $\eta>0$ and label all datapoints $x_i\in\Zsafe$ as a boundary point whose cardinality of  reverse $k$-nearest neighbors is below $\eta$.

Algorithm \ref{alg:mr-peanut} summarizes the boundary point detection algorithm. We first compute the pairwise distances between each of the $N_1$ safe datapoints in $Z_\text{safe}$ (line 1). The result is a symmetric $N_1\times N_1$ matrix $M$ where the element at position $(i,j)$ represents the pairwise distance between the states $x_i$ and $x_j$, i.e., $M_{ij}:=\|x_i-x_j\|$.  Next, we calculate the $k$-nearest neighbors of each $x_i$, denoted by $kNN_i$, as the set of indices corresponding to the $k$ smallest column entries in the $i$th row of $M$ (line 2). We calculate the reverse $k$-nearest neighbors of each $x_i$ as $RkNN_i:=|\{x_j\in \Zsafe|x_i\in kNN_j\}|$ (line 3).  We then threshold each $RkNN_i$ by $\eta$ (line 4), i.e., $z_i:=\mathbbm{1}(RkNN_i \leq \eta)$ where $\mathbbm{1}$ is the indicator function. We obtain a tuple $(z_1,\hdots,z_{N_1})\in\mathbb{R}^{N_1}$, where the indices $i$ corresponding to states $x_i=1$  are  boundary points. 

\begin{algorithm}
	\centering
	\begin{algorithmic}[1]
		\Statex \textbf{Input: } Set of safe states $Z_\text{safe}$, number of nearest neighbors $k$, neighbor threshold $\eta > 0$
		\Statex \textbf{Output: } Set of boundary points, i.e., the set of as unsafe labeled states $Z_\mathcal{N}$ 
		\State $M \gets compute\_pairwise\_dists(Z_\text{safe})$ \quad\# compute pairwise distances between elements in $Z_\text{safe}$
		\State $kNN_i \gets comp\_k\_near\_neighbors(M)$ \quad\# compute $k$-nearest neighbors of $x_i\in\Zsafe$ 
		\State $RkNN_i \gets comp\_reverse\_k\_near\_neighbors(kNN)$ \quad\# compute reverse $k$-nearest neighbors of $x_i\in\Zsafe$ 
		\State $z_i \gets \mathbbm{1}(RkNN_i \leq \eta)$ \quad\# threshold $RkNN_i$ by $\eta$
	\end{algorithmic}
	\caption{Boundary Point Detection - \textsc{BPD}($Z_\text{safe}$,$k$,$\eta$)}
	\label{alg:mr-peanut}
\end{algorithm}

We have not yet specified the paramters $\epsilon$ and $\epsilon_\mathcal{N}$  to be able to check the constraints \eqref{thm:0_eq}, \eqref{eq:propp}, and \eqref{thmmmm}. While the value of $\epsilon$ merely defines the set of admissible states $\mathcal{D}$ and determines the size of the safe set $\mathcal{C}$ as discussed in Section \ref{sec:optim}.\ref{sec:conditions}, the value of $\epsilon_\mathcal{N}$ is important as the set of unsafe labeled states $\mathcal{N}$ should fully enclose $\mathcal{D}$. This imposes an implicit lower bound on $\epsilon_\mathcal{N}$  to guarantee safety. Therefore, one can artificially sample additional datapoints in proximity of $Z_\mathcal{N}$ and add these to the set $Z_\mathcal{N}$.  One  way to get an estimate of  $\epsilon_\mathcal{N}$ is to calculate the distance of each datapoint to the closest datapoint in $Z_\mathcal{N}$, respectively. Then, taking the maximum or an average over these values gives a good estimate of  $\epsilon_\mathcal{N}$.

Finally, we discuss what behavior expert demonstrations in $\Zdynamics$ should exhibit. 
We focus on the ROCBF constraint \eqref{eq:cbf_const_}, which must be verified to hold for some $u \in \U$, by using the expert demonstration $(y_i,t_i,u_i)$ in \eqref{eq:opt_3}.  The more transverse the vector field of the input dynamics $\langle \hat{G}(\hat{X}(y_i),t_i), u_i \rangle$ is to the level sets of the function $h(\hat{X}(y_i))$ {(i.e., the more parallel it is to the inward pointing normal $\nabla h(\hat{X}(y_i))$)}, the larger the inner-product term in constraint \eqref{eq:opt_3} will be without increasing the Lipschitz constant of $h(x)$.  This means that the expert demonstrations should demonstrate how to move away from the  unsafe labeled set.

\subsection{Solving the Constrained Optimization Problem}
\label{sec:compute}

Some remarks are in order with respect to the optimization problem \eqref{eq:opt}. If the extended class $\mathcal{K}$ function $\alpha$ is linear and  $\calH$ is parameterized as
$\calH := \{\langle \phi(x), \theta \rangle| \theta \in \Theta \}$
where $\Theta\subseteq\mathbb{R}^l$ is a convex set and $\phi:\mathbb{R}^n\to\mathbb{R}^l$ is a known basis function, then the optimization problem~\eqref{eq:opt} is convex. Note here in particular that $\|\nabla h(x)\|_\star$ is convex in $\theta$ since 1) $\nabla h(x)$ is linear in $\theta$, 2)  norms are convex functions, and 3) composition of a convex with a linear function preserves convexity. We remark that  rich 
function classes such as infinite dimensional reproducing kernel Hilbert spaces can be approximated to arbitrary accuracy
with such an $\calH$ \cite{rahimi2008random}. 

In the more general case when $\calH := \{ h(x;\theta) |  \theta \in \Theta \}$,
such as when $h(x;\theta)$ is a deep neural network or when $\alpha$ is a general nonlinear function, the optimization problem~\eqref{eq:opt}
is nonconvex. Due to the computational complexity
of general nonlinear constrained programming, we propose an unconstrained
relaxation of the optimization problem~\eqref{eq:opt}. 
Our unconstrained relaxation results in the optimization problem:
\begin{subequations}\label{eq:opt_relaxed}
	\begin{align}
		\max_{\lambda_{\mathrm{s}},\lambda_{\mathrm{u}},\lambda_{\mathrm{d}}}&\min_{\theta \in \Theta}  \norm{\theta}^2 + \lambda_{\mathrm{s}} \sum_{x_i \in \underline{Z}_\text{safe}} \Big[\gammasafe - h(x_i;\theta)\Big]_+ \\
		&+ \lambda_{\mathrm{u}} \sum_{x_i \in \ZN} \Big[h(x_i;\theta) + \gammaunsafe\Big]_+ + \lambda_{\mathrm{d}} \sum_{(y_i,t_i,u_i) \in \Zdynamics} \Big[\gammadynamics - q(y_i,t_i,u_i;\theta)\Big]_+  
	\end{align}
\end{subequations}
where $[r]_+ := \max\{r, 0\}$ for $r \in \RR$ and where the function $q(u,y,t;\theta)$ is as in \eqref{eq_q_def} but now defined via $h(x;\theta)$. The positive parameters 
$\lambda_{\mathrm{s}}$, $\lambda_{\mathrm{u}}$, and $\lambda_{\mathrm{d}}$ are dual variables. While the unconstrained optimization problem \eqref{eq:opt_relaxed} is in general a nonconvex optimization problem, it can be solved efficiently in practice by iteratively solving the outer and inner optimization problems with respect to $(\lambda_{\mathrm{s}},\lambda_{\mathrm{u}},\lambda_{\mathrm{d}})$ and $\theta$, respectively, with stochastic first-order gradient methods
such as Adam or stochastic gradient descent \cite{chamon2020probably}.


\subsection{Hyperparameters and Lipschitz Constant Estimation}
\label{sec:Lipschitz_constants}

We treat $(\alpha,\gamma_\text{safe},\gamma_\text{unsafe},\gamma_\text{dyn},L_h,\overline{\text{Lip}}_{B},k,\eta)$ as hyperparameters and bootstrap over them. This is a common technique in machine learning and usually done via grid search. Due to the nonconvexity of the optimization problem, one may not be able to satisfy all constraints in  \eqref{eq:opt_1}-\eqref{eq:opt_3},  \eqref{thm:0_eq}, \eqref{eq:propp}, \eqref{thmmmm}. We hence terminate the while loop in line 9 of Algorithm \ref{alg:overview} when a satisfactory empirical behavior is achieved.

The conditions in equations \eqref{thm:0_eq}, \eqref{eq:propp}, and \eqref{thmmmm} depend on  Lipschitz constants of the functions $h$ and $q$. Since we assume that $h$ is twice continuously differentiable and we restrict ourselves to a compact domain $\N \cup \D$, we have that
$h$ and $\nabla h$ are both uniformly Lipschitz over $\N \cup \D$.
In \cite{robey2020learning}, we showed two examples of $\calH$ (we considered DNNs and functions parametrized by random Fourier features) where
an upper bound on the
Lipschitz constants of functions $h\in\calH$ and its gradient $\nabla h(x)$ can be efficiently estimated, and we refer the interested reader to \cite{robey2020learning}.

\section{Simulations}
\label{sec:sim}

\begin{figure*}
	\begin{subfigure}{0.32\textwidth}
		\centering
		\includegraphics[scale=0.3]{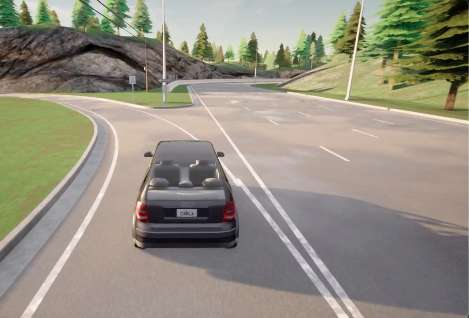}
		\caption{Training course.}
		\label{fig:CARLA1111}
	\end{subfigure}
	\begin{subfigure}{0.32\textwidth}
		\centering
		\includegraphics[scale=0.1035]{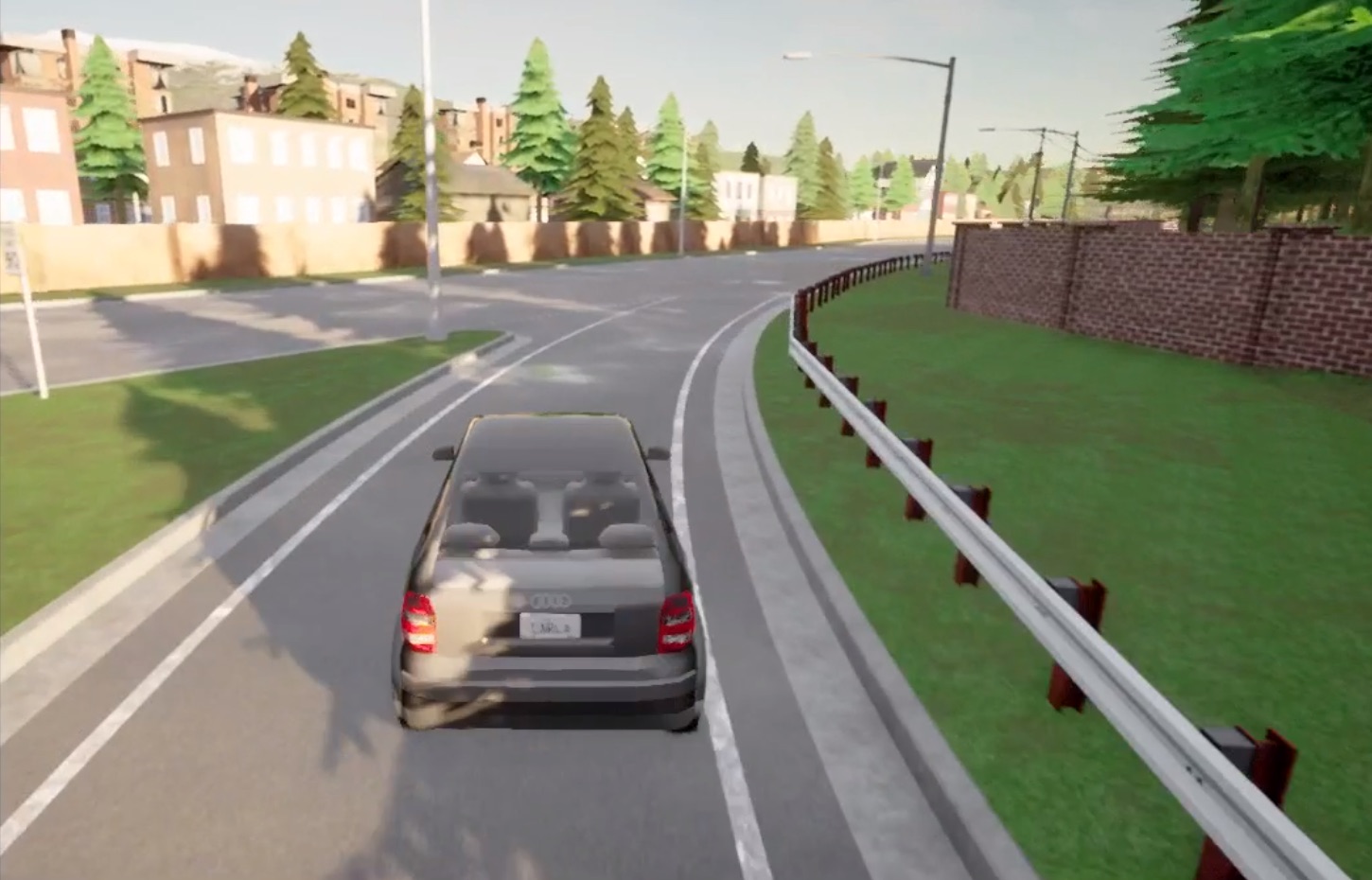}
		\caption{Test course.}
		\label{fig:CARLA2222}
	\end{subfigure}
	\begin{subfigure}{0.32\textwidth}
		\centering
		\includegraphics[scale=0.2]{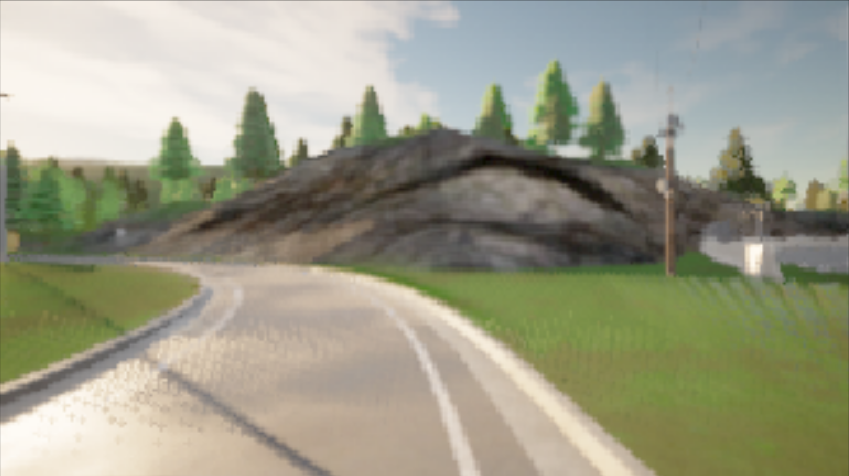}
		\caption{Downsampled   camera image.}
		\label{fig:dashboard}
	\end{subfigure}
	\caption{Simulation environment in CARLA. The cars are tracking desired reference paths on different courses. Left: The \textbf{training course} from which training data, during a left turn, was generated to train and test the ROCBF. Middle: An unknown \textbf{test course} on which the learned ROCBF is tested.  Right: Downsampled RGB dashboard camera image. }
	\label{fig:CARLA_}
\end{figure*}

We construct a safe ROCBF within the autonomous driving simulator CARLA \cite{dosovitskiy2017carla} for a car driving on a road by using camera images, see Fig. \ref{fig:CARLA_}.  In particular, our goal is to learn a ROCBF for the lateral control of the car, i.e., a lane keeping controller, while we use a built-in controller for longitudinal control. Lane keeping in CARLA is achieved by tracking a set of predefined waypoints. The control problem at hand is challenging  which makes it difficult to satisfy all constraints in equations \eqref{eq:opt_1}-\eqref{eq:opt_3} and  \eqref{thm:0_eq}, \eqref{eq:propp}, \eqref{thmmmm}. As described in Section \ref{sec:Lipschitz_constants}, we search over the hyperparameters of Algorithm \ref{alg:overview} until satisfactory behavior is achieved. The code for our simulations and videos of the car under the learned safe ROCBFs are available at 
\begin{center}
	\href{https://github.com/unstable-zeros/learning-rocbfs}{https://github.com/unstable-zeros/learning-rocbfs}.
\end{center}


As we have no direct access to the system dynamics of the car, we  identify a system model. The model for longitudinal control is estimated from data and consists of the velocity $v$ of the car and the integrator state $d$ of the PID. The identified longitudinal model of the car is 
\begin{align*}
	\dot{v}&=-1.0954v-0.007v^2-0.1521d+3.7387\\
	\dot{d}&=3.6v-20.
\end{align*}
For the lateral control of the car, we consider a bicycle model
\begin{align*}
	\dot{p}_x&=v\cos(\theta),\\
	\dot{p}_y&=v\sin(\theta),\\
	\dot{\theta}&=v/L\tan(\delta),
\end{align*}
where $p_x$ and $p_y$ denote the position in a global coordinate frame, $\theta$ is the heading angle with respect to a global coordinate frame, and $L:=2.51$ is the distance between the front and the rear axles of the car. The control input  is the steering angle $\delta$ that we design such that the car tracks  waypoints provided by CARLA. Treating $u:=\tan(\delta)$ as the  control input yields a control affine system.

To be able to learn a ROCBF  in a data efficient manner, we convert the above lateral model (defined in a global coordinate frame) into a local coordinate frame. We do so relatively to the waypoints that the car has to follow.  We consider the cross-track error $c_e$ of the car. In particular, let $wp_1$ be the waypoint that is closest to the car and let $wp_2$ be the waypoint proceeding $wp_1$. Then the cross-track error is defined as $c_e:=\|w\|\sin(\theta_w)$ where $w\in\mathbb{R}^2$ is the vector pointing from $wp_1$ to the car and $\theta_w$ is the angle between $w$ and the vector pointing from $wp_1$ to $wp_2$. We further consider the error angle $\theta_e:=\theta-\theta_t$ where $\theta_t$ is the angle between the vector  pointing from $wp_1$ to $wp_2$ and the global coordinate frame.  The simplified local model is 
\begin{align*}
	\dot{c}_e &=v\sin(\theta_e),\\
	\dot{\theta}_e&=v/2.51u-\dot{\theta}_t.
\end{align*}

In summary, we have the state $x:=\begin{bmatrix} v & d & c_e & \theta_e \end{bmatrix}^T$ and the control input $u:=\tan(\delta)$ as well as the external input,  given during runtime, of $\dot{\theta}_t$. Consequently, let
\begin{align*}
	\hat{F}:=\begin{bmatrix}
		-1.095v-0.007v^2-0.152d+3.74\\
		3.6v-20\\
		v\sin(\theta_e)\\
		-\dot{\theta}_t
	\end{bmatrix}
	\hat{G}:=\begin{bmatrix}
		0\\
		0\\
		0\\
		v/2.51
	\end{bmatrix}
\end{align*}
along with estimated error bounds $\Delta_F(x,t):=0.1$ and $\Delta_G(x,t)=0.1$ (calculated from simulations).

For collecting safe expert demonstrations $Z_\text{dyn}$, we use an ``expert'' PID controller $u(x)$ that uses full state knowledge of $x$. Throughout this section, we use the parameters $\alpha(r):=r$, $\gamma_\text{safe}:=\gamma_\text{unsafe}:=0.05$, and $\gamma_\text{dyn}:=0.01$ to train safe ROCBFs $h(x)$. For the boundary point detection algorithm in Algorithm~\ref{alg:mr-peanut}, we select $k:=200$ and $\eta$ such that $40$ percent of the points in $Z_\text{safe}$ are labeled as boundary points.

\subsection{State-based ROCBF}
\label{sec:state-based}
\begin{figure*}
	\begin{subfigure}{0.32\textwidth}
		\centering
		\includegraphics[scale=0.1517]{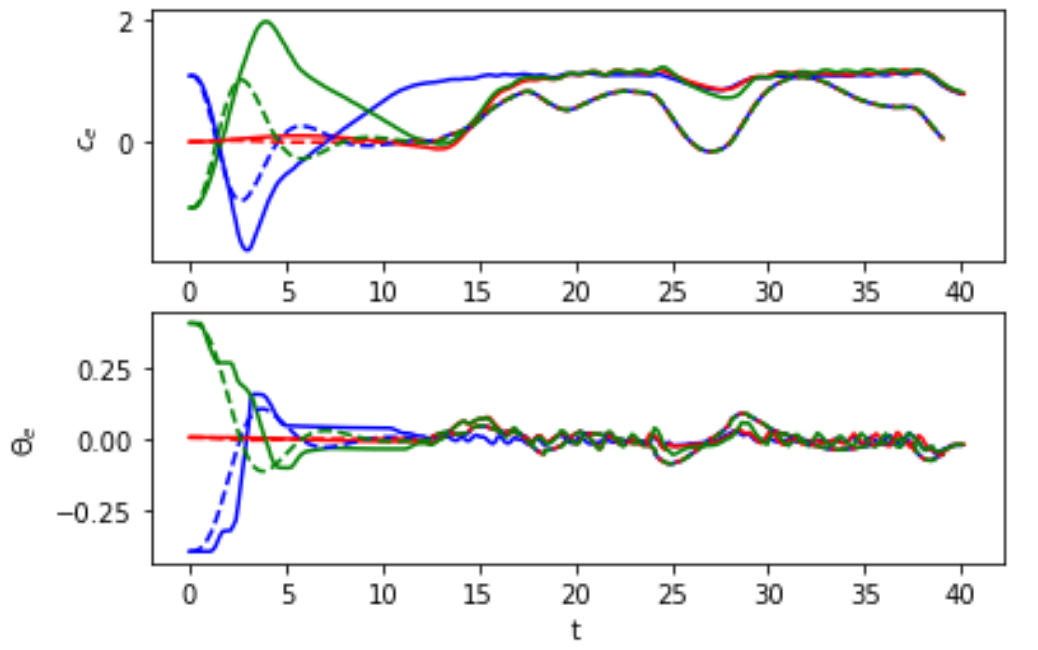}
		\caption{Trajectories $c_e(t)$ and $\theta_e(t)$ for three randomly chosen initial conditions. Solid lines correspond to the learned ROCBF,  dashed lines correspond to the expert (PID). }
		\label{fig:traj1}
	\end{subfigure}
	\begin{subfigure}{0.32\textwidth}
		\centering
		\includegraphics[scale=0.17]{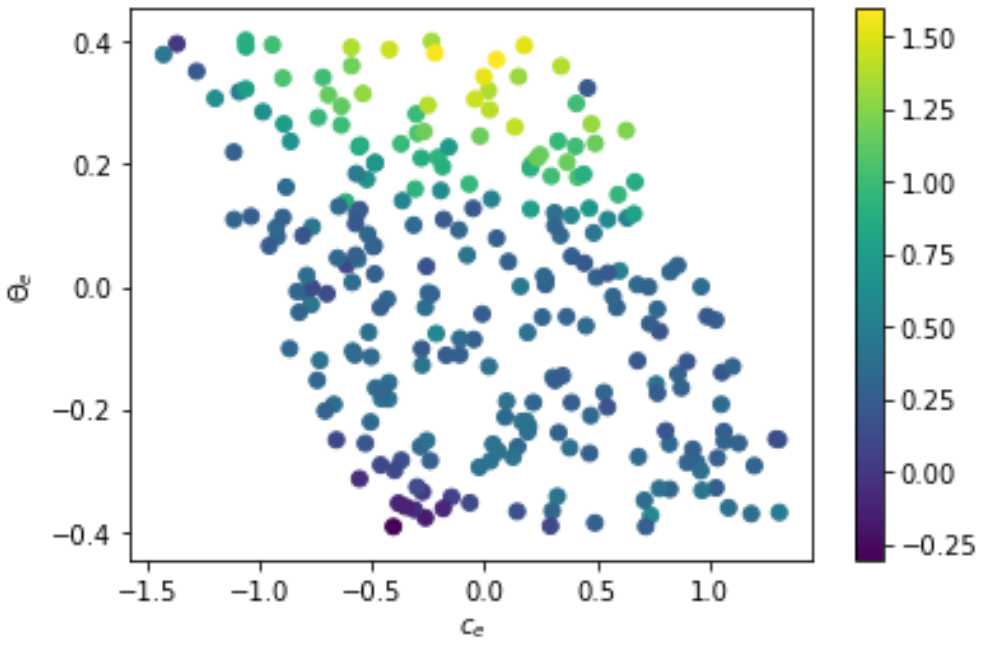}
		\caption{Shown are 250 different initial conditions on the \textbf{training course}. The color legend encodes $\max_t|c_e^{ROCBF}(t)|-\max_t|c_e^{Exp}(t)|$.}
		\label{fig:traj2}
	\end{subfigure}
	\begin{subfigure}{0.32\textwidth}
		\centering
		\includegraphics[scale=0.17]{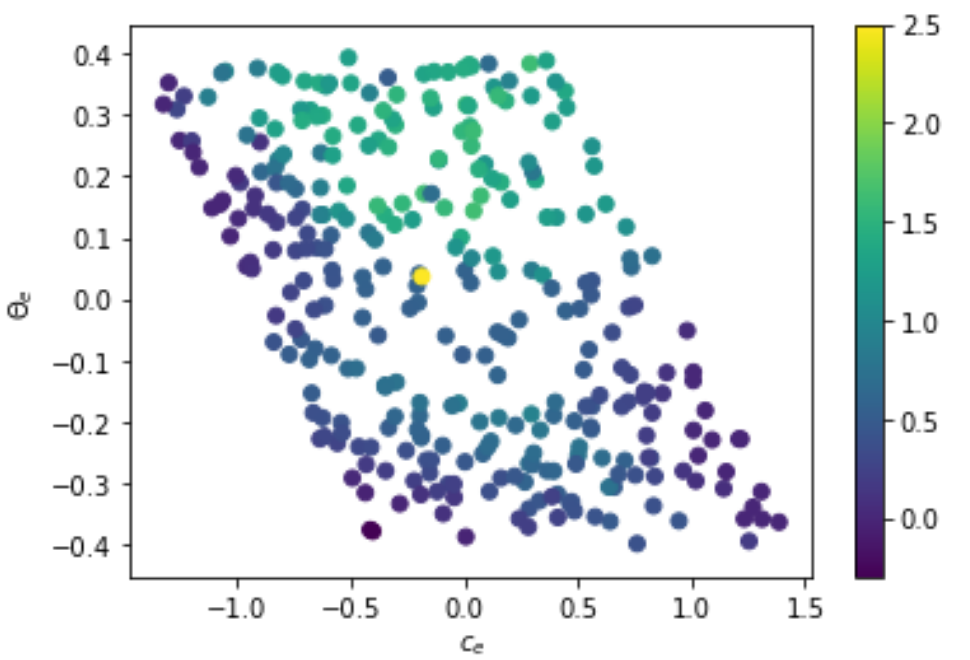}
		\caption{Shown are 300 different initial conditions on the \textbf{test course}. The color legend encodes $\max_t|c_e^{ROCBF}(t)|-\max_t|c_e^{Exp}(t)|$.}
		\label{fig:traj3}
	\end{subfigure}
	\caption{State-based ROCBF controller.}
	\label{fig:groundtruth}
\end{figure*}
We first learn a ROCBF controller in the case that the state $x$ is perfectly known, i.e., the model of the output measurement map is such that $\hat{X}(y)=X(y)=x$ and the error is $\Delta_X(y):=0$.  The trained ROCBF $h(x)$ is  a two-layer DNN with 32 and 16 neurons per layer. 

The safety controller applied to the car is then obtained as the solution of the convex  optimization problem $\min_{u\in \mathcal{U}} \|u\|$ subject to the constraint $q(u,y,t)\ge 0$. In Fig. \ref{fig:traj1},  example trajectories of $c_e(t)$ and $\theta_e(t)$ under this controller are shown. Solid lines indicate the learned ROCBF controller, while dashed lines indicate the expert PID controller for comparison. Colors in both subplots match the corresponding trajectories. The initial conditions of $d(0)$ and $v(0)$ are set to zero in all cases here, similar to all other plots in the remainder. Fig. \ref{fig:traj2} shows different initial conditions $c_e(0)$ and $\theta_e(0)$  and how the ROCBF controller performs relatively to the expert PID controller on the training course. In particular, each point in the plot indicates an initial condition from which system trajectories under both the ROCBF and expert PID controller are collected. The color map  shows 
\begin{align*}
	\max_t|c_e^{ROCBF}(t)|-\max_t|c_e^{Exp}(t)|
\end{align*}
where $c_e^{ROCBF}(t)$ and $c_e^{Exp}(t)$ denote the cross-track errors under the ROCBF and expert PID controllers, respectively.   Fig. \ref{fig:traj3} shows the same plot, but for the test course from which no data has been collected to train the ROCBF. In this plot, one ROCBF trajectory resulted in a collision as detected by CARLA. We assign by default a value of $2.5$ in case of a collision (see the yellow point in Fig. \ref{fig:traj3}).

\subsection{Perception-based ROCBF}
\begin{figure*}
	\begin{subfigure}{0.32\textwidth}
		\centering
		\includegraphics[scale=0.28]{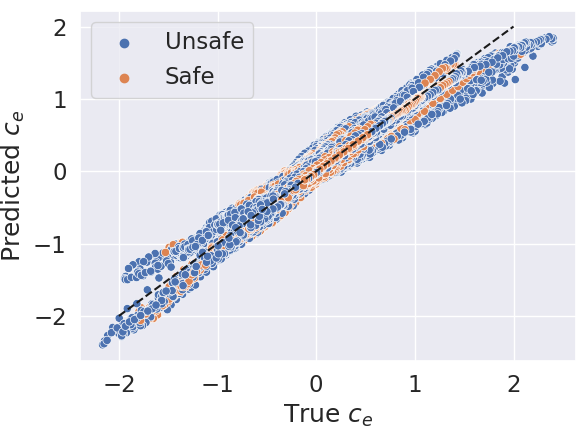}
		\caption{Performance of the perception map $\hat{X}$ on training data. Blue and orange are boundary and non-boundary points from Algorithm \ref{alg:mr-peanut}.}
		\label{fig:resnet}
	\end{subfigure}
	\begin{subfigure}{0.32\textwidth}
		\centering
		\includegraphics[scale=0.35]{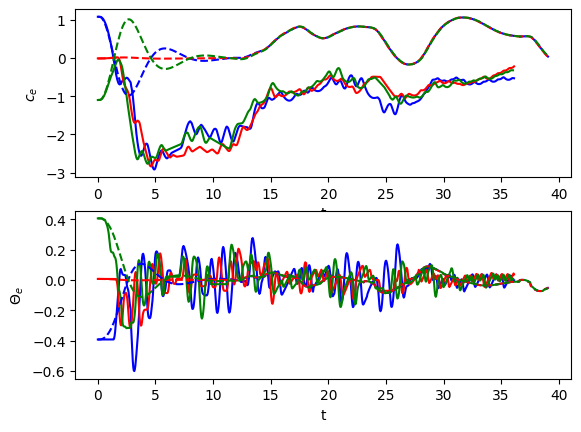}
		\caption{Trajectories $c_e(t)$ and $\theta_e(t)$ for randomly chosen initial conditions. Solid lines correspond to the learned ROCBF,  dashed lines correspond to the expert (PID). }
		\label{fig:traj1_}
	\end{subfigure}
	\begin{subfigure}{0.32\textwidth}
		\centering
		\includegraphics[scale=0.4]{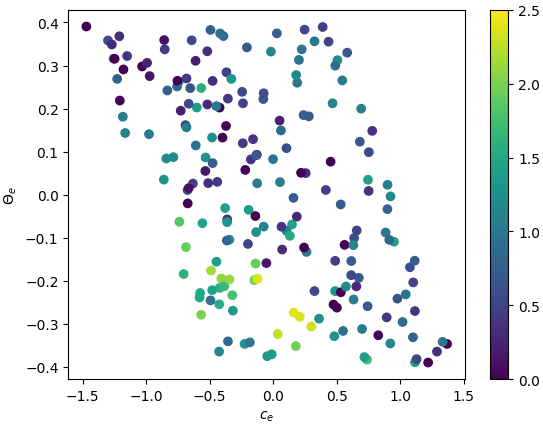}
		\caption{Shown are 200 different initial conditions on the \textbf{training course}. The color legend encodes $\max_t|c_e^{ROCBF}(t)|-\max_t|c_e^{Exp}(t)|$.}
		\label{fig:traj2_}
	\end{subfigure}
	\caption{Perception-based ROCBF controller.}
	\label{fig:perception}
\end{figure*}


We next learn a ROCBF in the case that $y$ corresponds to images taken from an RGB camera mounted to the dashboard of the car. To train a perception map $\hat{X}$, we have resized the images as shown in Fig. \ref{fig:dashboard}. We assume knowledge of $\theta_e$, $v$, and $d$, while we estimate $c_e$ from $y$, i.e., $x:=\begin{bmatrix} v & d & \hat{X}(y) & \theta_e \end{bmatrix}^T$. The architecture of $\hat{X}$ is a Resnet18, i.e., a convolutional neural network  with 18 layers. Its performance on  training data within operation range $c_e\in [-2,2]$ is shown in Fig. \ref{fig:resnet}. Based on this plot, we set $\Delta_X(y):=0.5$ to account for estimation errors within this range. We remark that we observed larger estimation errors outside this range. However, larger $\Delta_X(y)$ resulted in learning infeasible ROCBFs. We additionally selected the hyperparameter $\overline{\text{Lip}}_B(y,t,u):=\overline{\text{Lip}}_1+\overline{\text{Lip}}_2\|u\|$. We achieved the best results by using $\overline{\text{Lip}}_1=\overline{\text{Lip}}_2:=0.1$ during testing, while using the norm of the partial derivatives of $\langle\nabla \cbf(\xx), \hat{F}(x,t)\rangle+\alpha(\cbf(\xx))-\|\nabla \cbf(\xx)\|_2\Delta_F(x,t)$ and $\langle\nabla \cbf(\xx), \hat{G}(x,t)\rangle-\|\nabla \|_2\Delta_G(x,t)$, respectively, during training. The trained ROCBF $h(x)$ is again a two-layer DNN with 32 and 16 neurons per layer. Figs. \ref{fig:traj1_}-\ref{fig:traj2_} show the same plots as in the previous section and evaluate the ROCBF relatively to the expert PID controller. Importantly, note here that the expert PID controller uses state knowledge, while the ROCBF uses RGB images from the dashboard camera as inputs so that it is no surprise that the relative gap between these two becomes larger, as shown in Figs. \ref{fig:traj1_}-\ref{fig:traj2_}. 

\begin{figure}
	\centering
	\includegraphics[scale=0.3]{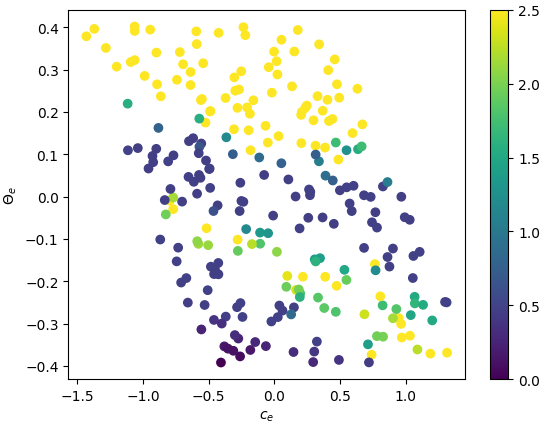}
	\caption{Comparison with learned CBF from \cite{robey2020learning}.}
	\label{fig:peanut}
\end{figure}

We further performed a comparison with our prior work \cite{robey2020learning} where we learn CBFs which corresponds to the setting when $\Delta_F=\Delta_G=\Delta_X=0$. The result of the learned CBF is shown in Figure \ref{fig:peanut}. In direct comparison with Figure \ref{fig:traj2_}, we can see the benefit of learning ROCBFs.


\section{Conclusion and Summary}
\label{sec:summary}

In this paper, we have shown how safe control laws can be learned from expert demonstrations under system model and measurement map uncertainties. We first presented robust output control barrier functions (ROCBFs) as a means to enforce safety, which is here defined as the ability of a system to remain within a safe set using the notion of forward invariance. We then proposed an optimization problem to learn such ROCBFs from safe expert demonstrations, and presented verifiable conditions for the validity of the ROCBF. These conditions are stated in terms of the density of the data and on Lipschitz  and boundedness constants of the learned function as well as the models of the system dynamics and the  measurement map. We proposed an algorithmic implementation of our theoretical framework to learn ROCBFs in practice. Finally, our simulation studies show how to learn safe control laws from RGB camera images within the autonomous driving simulator CARLA.

\section*{Appendix}

\appendix

\section{Proof of Theorem \ref{them:1}}
Recall that $f(t):=F(x(t),t)$, $g(t):=G(x(t),t)$, $y(t):=Y(x(t))$ and $u(t):=U(y(t),t)$ according to \eqref{eq:f}-\eqref{eq:u} and we define for convenience
\begin{align*}
	\hat{f}(t)&:=\hat{F}(x(t),t)\\
	\hat{g}(t)&:=\hat{G}(x(t),t)\\
	\delta_F(t)&:=\Delta_F(x(t),t)\\
	\delta_G(t)&:=\Delta_G(x(t),t)\\
	\hat{x}(t)&:=\hat{X}(y(t)).
\end{align*}
Due to the chain rule and since $U(y,t)\in \mathcal{U}_\text{s}(y,t)$, note that each solution $x:\mathcal{I}\to\mathbb{R}^n$ to \eqref{eq:system} under $U(y,t)$ satisfies
\begin{align}\label{eq:thm_1}
	\begin{split}
		B(\hat{x}(t),t,u(t)) \ge \text{Lip}_B(y(t),t,u(t))\Delta_X(y(t)).
	\end{split}
\end{align}
Note now that the term $\text{Lip}_B(y(t),t,u(t))\Delta_X(y(t))$ in the right-hand side of \eqref{eq:thm_1} can be bounded as
\begin{align*}
	&\text{Lip}_B(y(t),t,u(t))\Delta_X(y(t))\\
	\overset{(a)}{\ge}&\text{Lip}_B(y(t),t,u(t))\|\hat{x}(t)-x(t)\|\\
	\overset{(b)}{\ge}& |B(\hat{x}(t),t,u(t))-B(x(t),t,u(t))|\\
	\ge &B(\hat{x}(t),t,u(t))-B(x(t),t,u(t)) 
\end{align*}
where $(a)$ follows since $x(t)\in\mathcal{X}(y(t))$ due to Assumption \ref{ass:2} and where $(b)$ simply follows since $\text{Lip}_B(y(t),t,u(t))$ is the local Lipschitz constant of the function $B(x,t,u)$  within the set $\mathcal{X}(y(t))$. From \eqref{eq:thm_1} and the definitions of the functions $B(x,t,u)$ and $\text{Lip}_B(y,t,u)$, it hence follows that $B(x(t),t,u(t)) \ge 0$. Note next that the following holds
		\begin{align}
			&B(x(t),t,u(t)) \ge 0 \nonumber\\
			\Leftrightarrow&\langle\nabla \cbf(x(t)),
			\hat{f}(t)+\hat{g}(t)u(t)\rangle+\alpha(\cbf(x(t)))\nonumber\\
			&\ge \|\nabla \cbf(x(t))\|_\star(\delta_F(t)+\delta_G(t)\|u(t)\|)
			\nonumber\\
			\overset{(c)}{\Rightarrow}& \langle\nabla \cbf(x(t)), f(t)+g(t)u(t)\rangle+\alpha(\cbf(x(t)))\ge 0\nonumber\\
			\Leftrightarrow& \dot{h}(x(t))\ge -\alpha(\cbf(x(t)))\label{eq:thm_2}
		\end{align}
		where the implication in $(c)$ follows since
		\begin{align*}
			&\|\nabla \cbf(x(t))\|_\star(\delta_F(t)+\delta_G(t)\|u(t)\|)\\
			\overset{(d)}{\ge}  &\|\nabla \cbf(x(t))\|_\star(\|\hat{f}(t)-f(t)\|+\vertiii{\hat{g}(t)-g(t)}\|u(t)\|) \\
			\overset{(e)}{\ge}  &\|\nabla \cbf(x(t))\|_\star(\|\hat{f}(t)-f(t)\|+\|(\hat{g}(t)-g(t))u(t))\|)\\
			\overset{(f)}{\ge} &\langle \nabla\cbf(x(t)) ,\hat{f}(t)+\hat{g}(t)u(t)\rangle -\langle \nabla\cbf(x(t)) ,{f}(t)+{g}(t)u(t)\rangle
		\end{align*}
		where the inequality $(d)$ follows since $f(t)\in \mathcal{F}(x(t),t)$ and $g(t)\in \mathcal{G}(x(t),t)$ due to Assumption \ref{ass:1} and where the inequality  $(e)$ follows by properties of the induced matrix norm $\vertiii{\cdot}$. The inequality $(f)$ follows  by application of Hölder's inequality. Consequently, by \eqref{eq:thm_2} it holds that $\dot{h}(x(t))\ge -\alpha(\cbf(x(t)))$ for all $t\in\mathcal{I}$.
		
		Next note that $\dot{v}(t)=-\alpha(v(t))$ with $v(0)\ge 0$ admits a unique solution $v(t)$ that is such that $v(t)\ge 0$ for all $t\ge 0$ \cite[Lemma 4.4]{Kha96}. Using the Comparison Lemma \cite[Lemma 3.4]{Kha96} and assuming that $h(x(0))\ge 0$, it follows that $h(x(t))\ge v(t)\ge 0$ for all $t\in\mathcal{I}$, i.e., $x(0)\in\mathcal{C}$ implies $x(t)\in\mathcal{C}$ for all $t\in \mathcal{I}$. Recall  that \eqref{eq:system} is defined on $X(\mathcal{Y})$ and that $\mathcal{Y}\supseteq Y(\mathcal{C})$ so that $X(\mathcal{Y})\supseteq \C$. Since $x\in \mathcal{C}$ for all $t\in\mathcal{I}$ and when $\mathcal{C}$ is compact, it follows by \cite[Theorem 3.3]{Kha96}\footnote{Note that the same result as in \cite[Theorem 3.3]{Kha96} holds when the system dynamics are continuous, but not locally Lipschitz continuous.} that $\mathcal{I}=[0,\infty)$, i.e., $\mathcal{C}$ is forward invariant under $U(y,t)$. 
		
		
		\section{Proof of Proposition \ref{thm:0}}
		Note  that, for any $x\in\N$, there exists a point $x_i \in \ZN$ satisfying $\|x-x_i\|\leq \epsilon_\mathcal{N}$ since $\ZN$ is an $\epsilon_\mathcal{N}$-net of $\N$. For any $x\in\N$, we now select such an $x_i \in \ZN$ for which  
		\begin{align*}
			&\cbf(\xx) =  \cbf(\xx)-\cbf(\xx_i) + \cbf(\xx_i)\overset{(a)}{\leq} |\cbf(\xx)-\cbf(\xx_i)| - \gammaunsafe \\
			&\overset{(b)}{\leq} \text{Lip}_h(x_i)\|x-x_i\| - \gammaunsafe \overset{(c)}{\leq} \text{Lip}_h(x_i)\epsilon_\mathcal{N} - \gammaunsafe \overset{(d)}{<} 0.
		\end{align*}
		Note that inequality $(a)$ follows from constraint \eqref{eq:opt_2}, while inequality $(b)$ follows by Lipschitz continuity. Inequality $(c)$ follows by the assumption of $\ZN$ being an $\epsilon_\mathcal{N}$-net of $\N$ and, finally, the strict inequality in $(d)$ follows due to \eqref{thm:0_eq}.
		
		\section{Proof of Proposition \ref{cor:1}}
		The  proof follows similarly to the proof of Proposition \ref{thm:0}. For any $x\in\underline{\mathcal{D}}$, we select an $x_i \in \underline{Z}_\text{safe}$ with $\|x-x_i\|\le {\epsilon}$ which is possible since the set $\underline{Z}_\text{safe}$ is an ${\epsilon}$-net of $\underline{\mathcal{D}}$. It follows that  
		\begin{align*}
			0= h(x_i)-h(x_i)&\overset{(a)}{\leq} h(x_i)-h(x)+h(x) - \gammasafe \\
			&\overset{(b)}{\leq} \text{Lip}_h(x_i)\|x-x_i\|+h(x) - \gammasafe \\
			&\overset{(c)}{\leq} \text{Lip}_h(x_i){\epsilon}+ h(x)- \gammasafe \overset{(d)}{\le} h(x).
		\end{align*}
		Note that inequality $(a)$ follows from constraint \eqref{eq:opt_1}, while inequality $(b)$ follows by Lipschitz continuity. Inequality $(c)$ follows by  $\underline{Z}_\text{safe}$ being an ${\epsilon}$-net of $\underline{\mathcal{D}}$ and, finally, the inequality in $(d)$ follows due to \eqref{eq:propp}.

		\section{Proof that $y$ components of $\Zdynamics$ form an $\bar{\epsilon}$-net of $\mathcal{Y}$}
		\label{sec:F}
		
		\begin{lemma}\label{lemmmmm}Let $\bar{\epsilon}:=\text{Lip}_Y(\epsilon+\bar{\Delta}_X)$ where $\text{Lip}_Y$ is the  Lipschitz constant of the function $Y$ within the set $\overline{\mathcal{D}}:=\mathcal{D}\oplus \mathcal{B}_{2\bar{\Delta}_X}(0)$ where $\bar{\Delta}_X:=\sup_{y\in \mathcal{Y}}\Delta_X(y)$. Then the $y$ components of $\Zdynamics$ form an $\bar{\epsilon}$-net of $\mathcal{Y}$. 
		\end{lemma}
		\vspace{-0.25cm}
		\emph{Proof:} For each $y\in \mathcal{Y}$, there exists  $(y_i,t_i,u_i)\in{Z}_\text{dyn}$  such that       $\|X(y)-\hat{X}(y_i)\|\le \epsilon$
		by  definition of $\mathcal{Y}$ as $\mathcal{Y}=Y(\mathcal{D})$ and since  the $y$ components of ${Z}_\text{dyn}$ transformed via $\hat{X}$ form an $\epsilon$-net of $\mathcal{D}$. By Assumption \ref{ass:2}, we also know that  $\|X(y_i)-\hat{X}(y_i)\|\le \bar{\Delta}_X$.  By Lipschitz continuity of $Y$, it  follows that 
		\begin{align*}
			\|y-y_i\|&=\|Y(X(y))-Y(X(y_i))\|\le \text{Lip}_Y\|X(y)-X(y_i)\|\\
			&\le \text{Lip}_Y(\|X(y)-\hat{X}(y_i)\|+\|\hat{X}(y_i)-X(y_i)\|)\\
			&\le \text{Lip}_Y(\epsilon+\bar{\Delta}_X)=:\bar{\epsilon}.
		\end{align*}
		Consequently, the $y$ components of ${Z}_\text{dyn}$ form an $\bar{\epsilon}$-net of $\mathcal{Y}$.
		
		\section{Proof of Proposition \ref{prop:2}}
		Note first that, for each $y\in\mathcal{Y}$,  there exists a pair $(y_i,t_i,u_i) \in Z_\text{dyn}$ satisfying $\|y-y_i\|\leq \bar{\epsilon}$ since the $y$ component of $Z_\text{dyn}$ form an $\bar{\epsilon}$-net of $\mathcal{Y}$ by Lemma \ref{lemmmmm}. For any pair $(y,t)\in\mathcal{Y}\times\mathbb{R}_{\ge 0}$, we now select such a pair $(y_i,t_i,u_i) \in Z_\text{dyn}$ satisfying $\|y-y_i\|\leq \bar{\epsilon}$ for which then
		\begin{align*}
			0&\overset{(a)}{\leq} q(y_i,t_i,u_i)-\gamma_\text{dyn}\\
			&\leq|q(y_i,t_i,u_i)-q(y,t_i,u_i)|+q(y,t_i,u_i)-\gamma_\text{dyn}\\
			&\overset{(b)}{\leq}\text{Lip}_q(y_i,t_i,u_i)\|y_i-y\|+q(y,t_i,u_i)-\gamma_\text{dyn}\\
			&\overset{(c)}{\leq}\text{Lip}_q(y_i,t_i,u_i)\bar{\epsilon}+q(y,t_i,u_i)-\gamma_\text{dyn}\\
			&\le \text{Lip}_q(y_i,t_i,u_i)\bar{\epsilon}+|q(y,t_i,u_i)-q(y,t,u_i)|\\
			&\hspace{0.32cm}+q(y,t,u_i)-\gamma_\text{dyn}\\
			&\overset{(d)}{\leq}\text{Lip}_q(y_i,t_i,u_i)\bar{\epsilon}+\text{Bnd}_q(y_i,u_i)+q(y,t,u_i)-\gamma_\text{dyn}\\
			&\overset{(e)}{\leq}q(y,t,u_i).
		\end{align*}
		Inequality $(a)$ follows from  constraint \eqref{eq:opt_3}. Inequality $(b)$ follows by  Lipschitz continuity, while inequality $(c)$ follows since the $y$ component of $Z_\text{dyn}$ is an $\bar{\epsilon}$-net of $\mathcal{Y}$. Inequality $(d)$ follows by the bound $\text{Bnd}_q(y_i,u_i)$ that bounds the function $q$ for all values of $t$. Inequality $(e)$ follows simply by \eqref{thmmmm}. Consequently, $q(y,t,u_i)\ge 0$ for all $(y,t)\in\mathcal{Y}\times\mathbb{R}_{\ge 0}$. 
		
		If now   ${\text{Lip}}_B(y,t,u)\le \overline{\text{Lip}}_B(y,t,u)$, as stated per assumption, it follows  that \eqref{eq:cbf_const_} holds and that $h(x)$ is a ROCBF.

\bibliographystyle{IEEEtran}
\bibliography{main}

\addtolength{\textheight}{-12cm}   

\end{document}